\documentclass[nofootinbib,groupedaddress,superscriptaddress,aps,prx,reprint,floatfix]{revtex4-2}

\usepackage{amsmath,amssymb,mathtools,bm}
\usepackage{amsthm}
\usepackage[shortlabels]{enumitem}
\usepackage{graphicx,dsfont}
\newcommand{\bb}{\begin{equation}\begin{aligned}\hspace{0pt}}
\newcommand{\ee}{\end{aligned}\end{equation}}
\newcommand{\id}{\mathds{1}}
\usepackage{microtype}
\usepackage{xcolor}
\usepackage{hyperref}
\hypersetup{colorlinks=true,citecolor=blue,urlcolor=blue,linkcolor=blue}
\usepackage{times,txfonts}

\newcommand{\Tr}{\operatorname{Tr}}
\newcommand{\scnorm}[1]{\left\lVert#1\right\rVert}
\newcommand{\scket}[1]{\lvert#1\rangle}
\newcommand{\scbra}[1]{\langle#1\rvert}
\newcommand{\chiG}{\chi_{\mathrm{G}}}
\newcommand{\sgn}{\operatorname{sgn}}
\newcommand{\diag}{\operatorname{diag}}
\newcommand{\ketbra}[1]{\scket{#1}\!\scbra{#1}}

\theoremstyle{plain}
\newtheorem{theorem}{Theorem}
\newtheorem{proposition}[theorem]{Proposition}
\newtheorem{lemma}[theorem]{Lemma}
\newtheorem{corollary}[theorem]{Corollary}
\newtheorem{fact}[theorem]{Fact}
\makeatletter
\def\thmhead@plain#1#2#3{\thmname{#1}\thmnumber{\@ifnotempty{#1}{ }\@upn{#2}}\thmnote{ {\the\thm@notefont#3}}}
\let\thmhead\thmhead@plain
\makeatother
\theoremstyle{remark}

\begin{document}

\title{Classical Capacity of Single-Mode Bosonic Gaussian Channels}

\author{Gerardo Adesso}
\email{gerardo.adesso@nottingham.ac.uk}
\affiliation{School of Mathematical Sciences and Centre for the Mathematics and Theoretical Physics of Quantum Non-Equilibrium Systems, University of Nottingham, University Park, Nottingham NG7 2RD, United Kingdom}

\author{Ludovico Lami}
\email{ludovico.lami@sns.it}
\affiliation{Scuola Normale Superiore, Piazza dei Cavalieri 7, 56126 Pisa, Italy}

\author{Vittorio Giovannetti}
\email{vittorio.giovannetti@sns.it}
\affiliation{NEST, Scuola Normale Superiore and Istituto Nanoscienze-CNR, Piazza dei Cavalieri 7, 56126 Pisa, Italy}

\date{15 October 2026}

\begin{abstract}
The classical capacity of a noisy optical channel is the maximum rate at which information can be transmitted reliably, given a constraint on the mean input energy. For the ubiquitous bosonic Gaussian channels, which include standard models of loss and noise in fibre and free-space optical links, determining this rate requires addressing whether entanglement between channel uses or non-Gaussian signals can outperform Gaussian encodings. More precisely, the question is whether the energy-constrained Holevo information is additive across arbitrarily many channel uses and coincides with its Gaussian counterpart. We answer these fundamental questions in the affirmative for every single-mode bosonic Gaussian channel at every input energy. Namely, we prove that independently prepared Gaussian states with Gaussian-distributed random displacements achieve the capacity, including in the low-energy regime where squeezing the signals competes with encoding information in their displacement. Our proof connects an entanglement inequality to a sharp, additive lower bound on output entropy. We then characterise communication above capacity: under a strict total photon-number cutoff, the probability of correct decoding decreases exponentially with the number of channel uses, and we derive explicit bounds on the decay exponent. When only the mean photon number is constrained, a simple energy-rescaling formula determines exactly the entire asymptotic rate--error trade-off: the capacity at mean photon number $N$ and allowed average decoding error $\varepsilon$ equals the vanishing-error capacity at mean photon number $N/(1-\varepsilon)$. Together, these results identify the ultimate limits of classical communication for some of the most fundamental models of optical communication.
\end{abstract}

\maketitle

\section{Introduction}
The information carried by an optical signal is limited by quantum fluctuations, environmental noise, and the energy available to the transmitter. Bosonic Gaussian channels describe these effects for a broad range of optical and microwave links~\cite{CavesDrummond1994,HolevoWerner2001,Weedbrook2012}. Their classical capacity asks for the largest communication rate achievable with vanishing decoding error, allowing arbitrary quantum signal states and collective measurements at the receiver~\cite{Holevo1998,SchumacherWestmoreland1997}. Even for a channel acting on a single mode, this is a many-mode optimisation: entanglement between successive inputs might increase the information transmitted per use. As Hastings famously showed by resolving in the negative the additivity conjectures in quantum information theory, such an advantage is possible for general quantum channels~\cite{Hastings2009}.

For phase-insensitive bosonic Gaussian channels, Gaussian minimum-output-entropy theorems resolve this difficulty and establish the optimality of coherent states with Gaussian-distributed random displacements encoding the message~\cite{Giovannetti2004,GGCH2014,GHG2015,MGH2014,Holevo2015RMS}. Phase-sensitive noise, however, introduces an additional trade-off: indeed, squeezing the signal to match the noise minimises its output entropy, but consumes photons that could otherwise carry a message.
As it turns out, this trade-off identifies a precise threshold in the available energy: above, both requirements can be met; below, the optimal Gaussian ensemble is forced to use a different squeezing and random displacements in only one quadrature~\cite{Schafer2013,Karpov2013,Schafer2016}. Correlated-noise models encounter the same obstruction in their phase-sensitive normal modes~\cite{Pilyavets2008,LPM2009,Schafer2009,Schafer2011,PLM2012,Caruso2014}.

Ji~\cite{Ji2026} recently established Gaussian optimality of the energy-constrained single-use Holevo information, using the affine entanglement inequality originally introduced by Adesso in~\cite[Theorem~1]{Adesso2026}. Here we resolve the optimisation over correlated inputs by applying the newly established additivity result in version 3 of that work~\cite[Theorem~2]{Adesso2026}. In particular, we solve the classical-capacity problem for \emph{every} single-mode bosonic Gaussian channel at every input energy: independently prepared Gaussian signals attain the unrestricted capacity, and neither non-Gaussian inputs nor correlations across channel uses can increase the rate.

The key idea is to use a lower bound on the local entropy of a (multimode) bipartite state in terms of appropriately chosen second moments, originally found in~\cite{Adesso2026} and here reported in Proposition~\ref{epr_prop}. By exploiting it, we derive Theorem~\ref{entropy_thm}, which gives an output-entropy lower bound for parallel uses of Gaussian channels in terms of the second moments of the input. This also determines the optimal allocation of a shared energy budget across different links. For passively diagonalisable correlated noise, the same result identifies the optimal signals in the normal-mode basis. For nondegenerate channels at positive energy, the same analysis also implies that the capacity-achieving average output is unique and quantitatively constrains the average output of any near-optimal ensemble.

The capacity marks the largest rate at which information can be transmitted reliably, but it does not by itself say what happens when one attempts to transmit above this rate. 
Under the same mean-energy constraint, we determine exactly how much the transmission rate can increase when a nonzero average decoding error is allowed. The transmitter can devote more energy to a chosen fraction of the messages while keeping the mean energy fixed, accepting errors on the remainder. By contrast, under the stronger requirement that every codeword obey a hard total photon-number cutoff, this possibility is excluded and the probability of correct decoding decreases exponentially at every rate above capacity. We establish both results at every input energy and extend them to the correlated-noise model considered here. These conclusions hold even when the transmitter is allowed arbitrary, possibly non-Gaussian and entangled signals, and the receiver arbitrary collective measurements.

The remainder of the paper is organised as follows. In
Section~\ref{channels_sec} we introduce the channel model and the
Gaussian single-use optimisation. Section~\ref{entropy_sec}
establishes the sharp output-entropy bound, which is then used in
Section~\ref{capacity_sec} to derive the exact classical capacity and
optimal energy allocation. Section~\ref{memory_sec} extends these
results to channels with correlated Gaussian noise.
Section~\ref{strong_converse_sec} first characterises the average outputs of optimal and near-optimal ensembles, then determines the exact rate--error trade-off under the mean-energy constraint and establishes a complementary exponential strong converse under a hard photon-number cutoff. We conclude in Section~\ref{discussion_sec}
with a discussion of the results and open questions. Technical
proofs and quantitative converse bounds are collected in the
appendices.

\section{Channels and Gaussian Holevo information}\label{channels_sec}

We consider unassisted classical communication without feedback over bosonic Gaussian channels, with a constraint on the mean input photon number. We set $\hbar = 1$ and write $x,p$ for the quadrature operators of a mode, so that $[x,p] = i\id$ and its annihilation operator is $a = (x + ip)/\sqrt{2}$. For a state $\rho$, let $R \coloneqq (x,p)^{\mathsf{T}}$ denote the quadrature vector, let
$d \coloneqq \Tr[\rho R]$ be its vector of first moments, with components
$d_j = \Tr[\rho R_j]$, and let $V$ be its covariance matrix, with entries
\bb
V_{jk} \coloneqq \Tr\Big[\rho\big\{R_j - d_j\id,R_k - d_k\id\big\}\Big] \, .
\label{covariance}
\ee
where $\{B,C\} \coloneqq BC + CB$. Thus the vacuum has covariance matrix $V = \id$~\cite{AdessoRagyLee2014,BUCCO}; our covariance matrices are twice those used in~\cite{Adesso2026}. The mean photon number is
\bb
\Tr[\rho a^\dag a] = \frac{\Tr[V] + 2|d|^2 - 2}{4} \, .
\label{energy}
\ee

For a channel $\Phi$ acting on $m$ input modes, let
$H_m \coloneqq \sum_{j = 1}^m a_j^\dag a_j$. An ensemble
$(p_i,\rho_i)_i$ with average $\bar\rho \coloneqq \sum_i p_i\rho_i$
has Holevo information~\cite{Holevo1998,SchumacherWestmoreland1997}
\bb
\chi\big(\Phi;(p_i,\rho_i)_i\big)
\coloneqq S\big(\Phi(\bar\rho)\big)
- \sum_i p_i S\big(\Phi(\rho_i)\big) \, ,
\label{ensemble_holevo}
\ee
where $S(\rho) \coloneqq -\Tr[\rho\ln\rho]$ is the von Neumann
entropy. We use natural logarithms throughout, so entropies and
communication rates are measured in nats; the figures are reported
in bits. For continuous ensembles, the sums above are replaced by
integrals.

Maximising the Holevo information over all ensembles
whose average input obeys the energy constraint defines the
energy-constrained Holevo information of the channel:
\bb
\chi(\Phi,E) \coloneqq \sup_{(p_i,\rho_i)_i:\ \Tr[\bar\rho H_m] \leq E}\chi\big(\Phi;(p_i,\rho_i)_i\big) \, .
\label{channel_holevo}
\ee
Here and below, the optimisation includes continuous ensembles with finite mean energy. The corresponding \emph{classical capacity}, measured per use of the entire channel $\Phi$, is
\bb
C(\Phi,E) = \lim_{n\to\infty}\frac{1}{n}\chi\big(\Phi^{\otimes n},nE\big) \, .
\label{capacity_regularisation}
\ee
This is the capacity for vanishing average decoding error~\cite{HolevoBook,HolevoShirokov2006}.

After removing an irrelevant output displacement, a single-mode Gaussian channel acts on first and second moments of an arbitrary input state as $d\mapsto Xd$ and $V\mapsto XVX^{\mathsf{T}} + Y$, where the real $2\times2$ matrices $X,Y$ satisfy $Y \geq 0$ and $\det Y \geq (1 - \det X)^2$~\cite{HolevoBook}.  We call a Gaussian unitary without displacements a \emph{symplectic unitary}; its action on quadratures is $U^\dag R U = SR$ for a real symplectic matrix $S$. If $\tau \coloneqq \det X \neq 0$ and $Y>0$, an input phase rotation and an output symplectic unitary bring the channel to the form specified by
\bb
X = \sqrt{k}\,\diag(1,\varsigma),\qquad Y = y\,\diag(\xi^{-1},\xi) \, ,
\label{canonical}
\ee
where $k \coloneqq |\tau|$, $\varsigma \coloneqq \sgn\tau$, $0<\xi \leq 1$, and $y>0$ satisfies $y \geq |1 - \tau|$. We refer to~\eqref{canonical} as the canonical form~\cite{Schafer2013}; the reduction leading to it is given in Appendix~\ref{canonical_app}. Because the required input rotation preserves photon number and the output unitary preserves the Holevo information, this reduction does not affect the communication problem under consideration. The form~\eqref{canonical} includes attenuators ($0<\tau<1$), amplifiers ($\tau>1$), phase conjugators ($\tau<0$), and additive-noise channels ($\tau=1$) with positive definite noise, while the phase-insensitive subclass is recovered for $\xi=1$.

In what follows, we shall work under the nondegeneracy assumptions $\det X\neq0$ and $Y>0$. The remaining cases are recovered by continuity in Appendix~\ref{continuity_app}, thus ensuring that our final capacity result holds for all single-mode bosonic Gaussian channels.

\subsection{Gaussian signal ensembles}

A Gaussian state $\rho$ is uniquely determined by its first moments and covariance matrix \cite{AdessoRagyLee2014,BUCCO}. For one mode its von Neumann entropy depends only on the symplectic eigenvalue $\mu = \sqrt{\det V} \geq 1$ and is given by \cite{HolevoWerner2001}
\bb
S(\rho)=G(\mu) \coloneqq \frac{\mu + 1}{2}\ln\frac{\mu + 1}{2} - \frac{\mu - 1}{2}\ln\frac{\mu - 1}{2} \, .
\label{gaussian_entropy}
\ee
In particular, $G(\mu)$ is the entropy of the thermal state with covariance matrix $V = \mu\id$, whose mean photon number is $(\mu - 1)/2$.

A Gaussian signal ensemble consists of a fixed Gaussian input state, also known as the seed, together with a Gaussian distribution of displacements applied to it. The displacement distribution may be supported on a single quadrature; for example, displacement only in $p$ leaves the $x$ displacement equal to zero for every signal. We denote by $\chiG(\Phi,E)$ the supremum in~\eqref{channel_holevo} restricted to such ensembles, and call it the \emph{energy-constrained Gaussian Holevo information}. Throughout, $\chiG(\Phi,E)$ refers to this single-use optimisation.

Fix a Gaussian channel in the canonical form~\eqref{canonical}. To calculate $\chiG(\Phi,N)$, we first fix the average input state $\bar\rho$ to be the centred Gaussian state with covariance matrix $\diag(z,w)$, where $z,w>0$ and $zw \geq 1$, and optimise over Gaussian pure-state ensembles whose average is exactly $\bar\rho$. Only after this optimisation do we vary $z,w$ under the energy constraint. The fact that we can take $\bar\rho$ to have diagonal covariance matrix is proved in~\cite[Sec.~III, after Eq.~(14)]{Schafer2016}. For the fixed average state, the first term of the Holevo information is
\bb
h(z,w) \coloneqq G\!\left(\sqrt{(kz + y/\xi)(kw + y\xi)}\right) .
\label{h}
\ee

For a fixed average state $\bar\rho$, maximising the Holevo information therefore amounts to minimising the average output entropy of the individual signal states. It suffices to use a pure Gaussian seed with Gaussian-distributed random displacements~\cite[Supplemental Material, Eqs.~(55)--(62)]{Schafer2013}.

A pure squeezed vacuum $\gamma_u$ has covariance matrix $V_u = \diag(u,1/u)$ and output entropy $h(u,1/u)$. To obtain the prescribed $\bar\rho$ from displaced copies of $\gamma_u$, the centred Gaussian distribution of displacements must have covariance matrix $[\diag(z,w) - V_u]/2$. This is a valid covariance matrix precisely when $1/w \leq u \leq z$. Since displacements leave output entropy unchanged, optimising the pure-state ensemble with this average input amounts to choosing the squeezed vacuum with the least output entropy in that interval:
\bb
\varepsilon(z,w)&\coloneqq \min_{1/w \leq u \leq z}h(u,1/u) = h(\hat u,1/\hat u),\\
\hat u &\coloneqq \min\!\left\{z,\max\!\left\{1/w,1/\xi\right\}\right\} \, .
\label{eps}
\ee
Here $\hat u$ is the point in $[1/w,z]$ closest to $1/\xi$: indeed, $u\mapsto h(u,1/u)$ decreases up to $1/\xi$ and increases thereafter. Without the restriction on the average state, $\gamma_{1/\xi}$ attains the minimum output entropy $S_{\min} = G(k + y)$~\cite{GHG2015,Schafer2013}. The function $\varepsilon$ is nonincreasing in either variable, because increasing $z$ or $w$ enlarges the interval over which the minimum in~\eqref{eps} is taken.

For this fixed $\bar\rho$, the optimal Gaussian ensemble has Holevo information $h(z,w) - \varepsilon(z,w)$. Optimising over the covariance matrix of $\bar\rho$ gives the standard single-use Gaussian optimisation~\cite[Eq.~(11)]{Schafer2013}, written here in terms of the two diagonal moments:
\bb
\chiG(\Phi,N) = \max_{\substack{z,w>0,\ zw \geq 1\\ z + w = 4N + 2}}\big\{h(z,w) - \varepsilon(z,w)\big\} \, .
\label{gaussian_cap}
\ee

There is a smallest input energy for which the ensemble can use the minimum-output-entropy state $\gamma_{1/\xi}$ and, through random displacements, make the average output covariance matrix proportional to the identity. According to~\cite[Corollary~2, Eq.~(12)]{Schafer2013} and~\cite[Eq.~(28)]{Schafer2016}, its value is
\bb
N_{\rm th} \coloneqq \frac12\left[\frac1\xi - 1 + \frac{y(\xi^{-1} - \xi)}{2k}\right] \, .
\label{threshold}
\ee
For input energies $N \geq N_{\rm th}$, the same ensemble attains both the largest entropy of the average output allowed by the energy budget and the smallest entropy of every individual output. Consequently, in this regime
\bb
\chiG(\Phi,N) = G\!\left(k(2N + 1) + \frac{y(\xi + \xi^{-1})}{2}\right) - S_{\min} \, .
\label{above_threshold}
\ee
When the input energy satisfies $0<N<N_{\rm th}$, with $N_{\rm th}$ defined in~\eqref{threshold}, it is not possible to use an input ensemble with seed $\gamma_{1/\xi}$ and a thermal average output state. The known optimum instead has $z = u_*<1/\xi$ and uses $\gamma_{u_*}$ displaced only in $p$~\cite[Sec.~III.B, Eqs.~(31), (43)--(46)]{Schafer2016}\cite[Eq.~(27)]{Karpov2013}. At $N = 0$ the only input is vacuum and the Holevo information vanishes. Appendix~\ref{gaussian_app} recalls this optimisation.

The distinction between a single-use optimisation and a regularised capacity matters here. The Gaussian capacity defined in~\cite[Eq.~(10)]{Schafer2013} is regularised, and~\cite[Corollary~2, Eq.~(12)]{Schafer2013} proves Gaussian additivity for input energies $N \geq N_{\rm th}$, with $N_{\rm th}$ defined in~\eqref{threshold}. It does not establish additivity when $0<N<N_{\rm th}$. By contrast,~\cite[Sec.~III, Eq.~(10)]{Schafer2016} calculates a single-use Gaussian optimisation. Our capacity result in Sec.~\ref{capacity_sec} removes this energy restriction and identifies the single-use Gaussian value with the unrestricted classical capacity at every input energy.

\section{A sharp output-entropy bound}\label{entropy_sec}

The Holevo information in~\eqref{ensemble_holevo} is the difference of two entropy terms. The entropy of the average output can be bounded from above by subadditivity and by replacing each marginal with the Gaussian state having the same moments; the main task is therefore to lower-bound the entropy of the individual outputs. We prove a bound that depends only on the input quadrature second moments and is tight for squeezed vacua. Keeping both moments is essential when the input energy is below the threshold in~\eqref{threshold}, because the optimal Gaussian seed then differs from the unconstrained minimum-output-entropy input.

\begin{theorem}[(Output entropy from second moments)]\label{entropy_thm}
Let $\Phi_1,\dots,\Phi_m$ be single-mode Gaussian channels in the canonical form~\eqref{canonical}. For every state $\rho$ of $m$ modes with finite mean energy, possibly entangled and non-Gaussian,
\bb
S\bigl((\Phi_1\otimes\cdots\otimes\Phi_m)(\rho)\bigr) \geq \sum_{j = 1}^m\varepsilon_j\bigl(2\langle x_j^2\rangle_\rho,2\langle p_j^2\rangle_\rho\bigr) \, .
\label{entropy}
\ee
Here $\varepsilon_j$ is the function $\varepsilon$ defined in~\eqref{eps}, evaluated with the parameters $k_j,y_j,\xi_j$ of the $j^{\,\text{th}}$ channel.
Equality holds for every product of squeezed vacua $\gamma_{u_j}$. Furthermore, for all $z_j,w_j>0$ such that $z_jw_j \geq 1$, the bound is tight under the corresponding upper bounds on the second moments:
\bb
\sum_{j = 1}^m\varepsilon_j(z_j,w_j) = \min\Big\{ &S\bigl((\Phi_1\otimes\cdots\otimes\Phi_m)(\rho)\bigr):\\
&2\langle x_j^2\rangle_\rho \leq z_j,\ 2\langle p_j^2\rangle_\rho \leq w_j\ \,\forall\, j\Big\}\, .
\label{entropy_tightness}
\ee
The minimum can be attained with all the moment constraints saturated.
\end{theorem}

\begin{proof}
Write $z_j = 2\Tr[\rho x_j^2]$ and $w_j = 2\Tr[\rho p_j^2]$. We first obtain lower bounds that are affine in these moments. The two auxiliary ingredients used in the proof, Proposition~\ref{epr_prop} and Lemma~\ref{witness_lemma}, are established in Appendix~\ref{entropy_app}. For each channel, choose a squeezed vacuum $\gamma_{u_j}$, with $u_j\neq1/\xi_j$, independently of the actual input $\rho$. Lemma~\ref{witness_lemma} constructs a dilation by a symplectic unitary on the input mode $A_j$ and two auxiliary modes $B_j,E_j$, initially in vacuum. After a local squeezing of the receiver $A_j$, the resulting unitary $\widetilde V_{j,u_j}$ sends $\gamma_{u_j}\otimes\ketbra{0}_{B_jE_j}$ to a two-mode squeezed vacuum on $A_jB_j$ and vacuum on $E_j$. Apply the product of these unitaries to $\rho_A$ and the auxiliary vacuum, then trace out $E = E_1\ldots E_m$:
\bb
\Omega_{AB} \coloneqq \Tr_E\Bigg[&\left(\bigotimes\nolimits_j\widetilde V_{j,u_j}\right)\big(\rho_A\otimes\ketbra{0}_{BE}\big)\left(\bigotimes\nolimits_j\widetilde V_{j,u_j}\right)^\dag\Bigg] \, .
\label{entropy_dilation}
\ee
Here $A = A_1\ldots A_m$, $B = B_1\ldots B_m$, and $|0\rangle_{BE}$ is the joint vacuum of the $2m$ auxiliary modes. The state $\Omega_{AB}$ has finite energy, but in general will be mixed and potentially non-Gaussian.
Its marginal on $A$ is unitarily equivalent to $(\Phi_1\otimes\cdots\otimes\Phi_m)(\rho)$, so it has the same entropy.

We apply the mixed-state form of the many-pair entropy inequality, Proposition~\ref{epr_prop}, to $\Omega_{AB}$. Lemma~\ref{witness_lemma} chooses the coefficients of each quadratic observable so that its expectation depends on only one input quadrature moment. Substituting its identity~\eqref{pullback} into~\eqref{epr}, for $u_j<1/\xi_j$ on every mode, gives
\bb
&S\big((\Phi_1\otimes\cdots\otimes\Phi_m)(\rho)\big)\\
&\quad\geq\sum_{j = 1}^m\left[h_j(u_j,1/u_j) + \left.\frac{d}{dv}h_j(v,1/v)\right|_{v = u_j}(z_j - u_j)\right] \, .
\label{main_tangent}
\ee
Here $h_j$ is the function $h$ defined in~\eqref{h}, evaluated with the parameters $k_j,y_j,\xi_j$ of the $j^{\,\text{th}}$ channel.
Each summand is the tangent to the squeezed-vacuum output entropy as a function of its position second moment. On any mode we may instead choose $u_j>1/\xi_j$, obtaining the tangent to $w\mapsto h_j(1/w,w)$ at $w = 1/u_j$, evaluated at $w_j$. The many-pair inequality is essential here: separate bounds on the marginal output entropies would not give this lower bound on their joint entropy.

We can now choose $u_j$ so that the affine corrections vanish. If $z_j<1/\xi_j$, take $u_j = z_j$; if $w_j<\xi_j$, take $u_j = 1/w_j$. These cases are mutually exclusive because $z_jw_j \geq 1$, and the resulting summand is exactly $\varepsilon_j(z_j,w_j)$. Otherwise both $z_j\geq1/\xi_j$ and $w_j\geq\xi_j$, so take $u_j\to1/\xi_j$. The seed entropy tends to $S_{\min,j}$ and its derivative tends to zero, giving the same conclusion. This derivative limit also holds when $k_j + y_j = 1$: the output symplectic eigenvalue minus one vanishes quadratically in $u_j - 1/\xi_j$, which compensates for the logarithmic divergence of the entropy derivative. Taking these scalar limits mode by mode proves~\eqref{entropy}.

For a product of squeezed vacua the interval in~\eqref{eps} collapses to a single point on each mode, so equality holds in~\eqref{entropy}. To prove~\eqref{entropy_tightness}, every feasible state satisfies
\bb
S\big((\Phi_1\otimes\cdots\otimes\Phi_m)(\rho)\big) &\geq \sum_j\varepsilon_j\big(2\langle x_j^2\rangle_\rho,2\langle p_j^2\rangle_\rho\big)\\
&\geq \sum_j\varepsilon_j(z_j,w_j),
\ee
where the second inequality uses that each $\varepsilon_j$ is nonincreasing in both arguments. Let $\hat u_j$ be the minimiser in~\eqref{eps} for the $j^{\,\text{th}}$ channel. The product $\bigotimes_j\gamma_{\hat u_j}$ attains this value and obeys the constraints. Moreover, displacing its $j^{\,\text{th}}$ mode to have first moments $\sqrt{(z_j - \hat u_j)/2}$ and $\sqrt{(w_j - 1/\hat u_j)/2}$ makes both constraints equalities, while leaving every output entropy unchanged.
\end{proof}

The same affine bounds also establish the joint convexity of $\varepsilon$, which will be used in the capacity argument. For a single channel they give
\bb
\!\!\!\!\varepsilon(z,w) = \max\Bigg\{&S_{\min},\\
&\sup_{0<u<1/\xi}\left[h(u,1/u) + \frac{d}{du}h(u,1/u)(z - u)\right],\\
&\sup_{0<v<\xi}\left[h(1/v,v) + \frac{d}{dv}h(1/v,v)(w - v)\right]\Bigg\} \, .
\label{eps_affine}
\ee
Indeed, for any feasible $q\in[1/w,z]$, the tangent bounds evaluated on $\gamma_q$, together with their nonpositive slopes and the inequalities $q\leq z$ and $1/q\leq w$, show that every affine function on the right-hand side is no larger than $h(q,1/q)$. Minimising over such $q$ therefore bounds the right-hand side by $\varepsilon(z,w)$. Equality is obtained by choosing $u=z$ when $z<1/\xi$, or $v=w$ when $w<\xi$; when both $z\geq1/\xi$ and $w\geq\xi$, the constant branch $S_{\min}$ is attained. Hence $\varepsilon$ is a supremum of affine functions and is jointly convex.

\section{Capacity and energy allocation}\label{capacity_sec}

We now use the output-entropy bound of Theorem~\ref{entropy_thm} to control the Holevo information of an arbitrary ensemble. The theorem provides a lower bound on the average entropy of the individual outputs, so it remains to bound from above the entropy of the average output. By subadditivity, this entropy is no larger than the sum of the marginal output entropies, and each marginal entropy can only increase when the state is replaced by the Gaussian state with the same moments. The determinant of the resulting covariance matrix is at most the product of its two diagonal second moments, yielding
\bb
S\big((\Phi_1\otimes\cdots\otimes\Phi_m)(\bar\rho)\big)
\leq \sum_{j=1}^m h_j(z_j,w_j).
\ee
Combining the two bounds gives the following result.

\begin{theorem}[(Moment-resolved Gaussian optimality)]\label{holevo_thm}
Let $\Phi_1,\dots,\Phi_m$ be single-mode Gaussian channels in the canonical form~\eqref{canonical}. For every ensemble $(p_i,\rho_i)_i$ of states of $m$ modes with finite mean energy, possibly entangled and non-Gaussian, whose average has second moments $z_j = 2\langle x_j^2\rangle$ and $w_j = 2\langle p_j^2\rangle$,
\begin{equation}
\chi\big(\Phi_1\otimes\cdots\otimes\Phi_m;(p_i,\rho_i)_i\big) \leq \sum_{j = 1}^m\big[h_j(z_j,w_j) - \varepsilon_j(z_j,w_j)\big].
\label{holevo}
\end{equation}
Here $h_j$ and $\varepsilon_j$ are the functions defined in~\eqref{h} and~\eqref{eps}, respectively, evaluated with the parameters $k_j,y_j,\xi_j$ of $\Phi_j$. For all admissible $(z_j,w_j)$, equality is attained by a product of Gaussian ensembles with these second moments.
\end{theorem}

\begin{proof}
Apply Theorem~\ref{entropy_thm} to every signal state in the ensemble. Since $\varepsilon_j$ is convex and second moments are affine in the state, Jensen's inequality bounds the average signal output entropy from below by $\sum_j\varepsilon_j(z_j,w_j)$. Subtracting this from the upper bound on the average output entropy proves~\eqref{holevo}. The product of the Gaussian displacement ensembles constructed after~\eqref{eps} attains equality, because its average covariance matrix is diagonal and every displaced signal has output entropy $\sum_j\varepsilon_j(z_j,w_j)$.
\end{proof}

Since $h_j$ is jointly concave and $\varepsilon_j$ is convex, the difference
$h_j(z,w)-\varepsilon_j(z,w)$ is jointly concave. Theorem~\ref{holevo_thm}
therefore reduces the ensemble optimisation to a concave optimisation over
the admissible diagonal second moments, with every feasible point attained
by a product Gaussian ensemble.
For an average input state with total mean photon number at most $E$, the second moments satisfy
$\sum_{j=1}^m (z_j+w_j)\leq 4E+2m$.
Using the reduction to canonical form, which preserves the input photon number, together with the continuity argument for the remaining degenerate channels given in Appendix~\ref{continuity_app}, we obtain the following.

\begin{corollary}[(Capacities and water-filling)]\label{capacity_cor}
For arbitrary single-mode Gaussian channels $\Phi_1,\dots,\Phi_m$,
\bb
\chi(\Phi_1\otimes\cdots\otimes\Phi_m,E)
=
\max_{\substack{N_j\geq0\\ \sum_jN_j\leq E}}
\sum_{j=1}^m \chiG(\Phi_j,N_j).
\label{waterfilling}
\ee
Here $\chiG(\Phi,N)$ is obtained by restricting the optimisation in
Eq.~\eqref{channel_holevo} to Gaussian signal ensembles. For each $\Phi$,
the function $N\mapsto\chiG(\Phi,N)$ is concave. In particular:
\begin{enumerate}[(i)]
\item for every single-mode Gaussian channel $\Phi$,
\[
\chi(\Phi^{\otimes n},nN)=n\,\chiG(\Phi,N),
\]
and hence
\[
C(\Phi,N)=\chiG(\Phi,N)
\]
at every input energy;

\item for any two finite products $\Phi$ and $\Psi$ of single-mode Gaussian channels,
\[
\chi(\Phi\otimes\Psi,E)
=
\max_{\substack{E_1,E_2\geq0\\E_1+E_2\leq E}}
\big[\chi(\Phi,E_1)+\chi(\Psi,E_2)\big];
\]

\item the classical capacity of
$\Phi_1\otimes\cdots\otimes\Phi_m$ at total energy $E$ per use is given by
the water-filling formula~\eqref{waterfilling}.
\end{enumerate}
\end{corollary}

\begin{figure*}[t]
\includegraphics[width=0.96\textwidth]{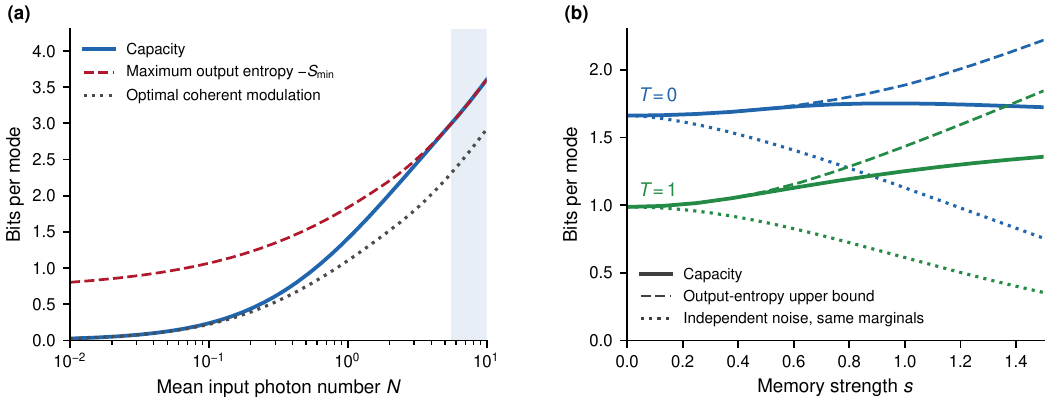}
\caption{{\bf (a)} Classical capacity of an attenuator with transmissivity $\eta = 0.7$ and squeezed thermal environment [$k = \eta$, $y = 3(1 - \eta)$, $\xi = e^{-2}$], as a function of the mean photon number $N$ (solid line). The dashed line is the upper bound $\max S(\text{output})-S_{\min}$. The light blue shaded region marks the above-threshold regime $N\geq N_{\rm th}\simeq5.5$, with $N_{\rm th}$ given by~\eqref{threshold}. In this regime, the optimal Gaussian ensemble attains the dashed upper bound, which therefore coincides with the exact capacity; below threshold, the bound is not tight. The dotted line is the best Holevo information achievable without input squeezing, using coherent state signals with Gaussian random displacements. {\bf (b)} Classical capacity per use of the lossy memory channel of Ref.~\cite{LPM2009} with $\eta = 0.7$, $N = 1$ and environment thermal occupancy $T \in \{0,1\}$, as a function of the memory parameter $s$ in the limit of many uses (solid line). The dashed line is the bound obtained from the maximum entropy of the average output minus the minimum output entropy of the normal modes, while the dotted line is the capacity of a memoryless channel with the same single-use marginals, i.e., thermal noise in the bulk limit with $T_{\rm eff} + \tfrac12 = (T + \tfrac12)I_0(2s)$, where $I_0$ is the modified Bessel function of the first kind.}
\label{capacity_fig}
\end{figure*}

Statement (i) completes the single-use result of Ji~\cite{Ji2026} and, together with Refs.~\cite{GGCH2014,GHG2015,Schafer2013}, determines the classical capacity of all single-mode Gaussian channels. When the input energy is below $N_{\rm th}$ in~\eqref{threshold}, the capacity is the Gaussian value determined by the transcendental optimisation of Refs.~\cite{Schafer2016,Karpov2013}, which is therefore not merely the best Gaussian rate. Figure~\ref{capacity_fig}(a) shows this for an attenuator with squeezed thermal noise. In the low-energy regime, the exact capacity lies well below the simpler bound obtained by maximising the entropy of the average output and subtracting the minimum output entropy; that bound becomes tight only for $N\geq N_{\rm th}$.

Statement (ii) is an energy-constrained strong additivity result within the class of single-mode Gaussian channels~\cite{MSW2004,Shor2004}: correlated encodings across different channel blocks provide no advantage, and the optimal Holevo information is obtained by dividing the total energy between the blocks and optimising them separately.

Statement (iii) turns this into an explicit energy-allocation rule for a collection of different Gaussian channels. Since each $\chiG(\Phi_j,N_j)$ is concave in $N_j$, the optimal allocation has the usual water-filling structure: among the channels that receive nonzero energy, the marginal gains $\partial_{N_j}\chiG(\Phi_j,N_j)$ are equalised. In the present quantum setting, however, the structure of the optimal signal within each channel also depends on the allocated energy~\cite{Schafer2011}. A channel may be left unused, operated below its threshold $N_{{\rm th},j}$ with a squeezed seed displaced in only one quadrature, or operated above threshold with displacements in both quadratures. Different carrier frequencies or mode costs are included by assigning positive costs $c_j$ and replacing the constraint by $\sum_j c_jN_j\leq E$, in which case the active modes equalise $\partial_{N_j}\chiG(\Phi_j,N_j)/c_j$.

\section{Communication with correlated noise}\label{memory_sec}

Correlated noise is a natural model for bosonic channels with memory~\cite{KretschmannWerner2005,Caruso2014}. In the quadrature ordering $(x_1,\ldots,x_n,p_1,\ldots,p_n)$, consider a Gaussian channel with transfer matrix
$X=\sqrt{k}(\id_n\oplus\varsigma\id_n)$ and noise covariance matrix
$Y=Y_x\oplus Y_p$, where $\varsigma=\pm1$. The matrices $Y_x$ and $Y_p$
describe correlations across channel uses, with no position--momentum cross
block. This class includes lossy channels with correlated environments
($\varsigma=1$, $k<1$)~\cite{Pilyavets2008,LPM2009,PLM2012}, additive
correlated noise ($k=1$)~\cite{Schafer2009,Schafer2011}, and the corresponding
amplifying and conjugating channels.

When $Y_x$ and $Y_p$ commute, the correlations can be diagonalised by the same passive transformation in both quadratures. The channel then decomposes into independent normal modes, so the capacity result of the previous section applies directly.

\begin{corollary}[(Memory channels)]\label{memory_cor}
If $Y_xY_p=Y_pY_x$, the $n$-use channel is equivalent, through a passive transformation that preserves the total photon number, to a product of single-mode channels with noise $\diag(y_{x,l},y_{p,l})$, where $y_{x,l}$ and $y_{p,l}$ are the eigenvalues of $Y_x$ and $Y_p$ on a common eigenvector. The capacity of this block, when independently repeated with energy $E$ per block, is given by Eq.~\eqref{waterfilling} over these normal modes.
\end{corollary}

Indeed, if $O$ is an orthogonal matrix diagonalising both $Y_x$ and $Y_p$, the transformation $x\mapsto Ox$, $p\mapsto Op$ is a passive linear-optical network acting identically on the input and output quadratures, and it commutes with the transfer matrix.

As a concrete example, consider the lossy memory channel of Lupo, Pilyavets and Mancini~\cite{LPM2009}, for which
$Y_x=(1-\eta)(2T+1)e^{s\Delta}$ and
$Y_p=(1-\eta)(2T+1)e^{-s\Delta}$, where $\Delta$ is the adjacency matrix of a chain. The normal modes are attenuators with squeezed thermal environments of covariance matrix
$(2T+1)\diag(e^{2s\cos\theta_l},e^{-2s\cos\theta_l})$, with
$\theta_l=\pi l/(n+1)$. Previous work obtained upper and lower bounds for this model, which coincide only at zero temperature and in part of the parameter space~\cite{LPM2009}. Corollary~\ref{memory_cor} instead determines the Holevo information of every finite block exactly. In the limit $n\to\infty$, the capacity per mode converges to
\bb
C_{\rm mem}(N)
\coloneqq
\sup_{\substack{N_\theta\geq0\\
\pi^{-1}\int_0^\pi N_\theta\,d\theta\leq N}}
\frac1\pi\int_0^\pi
\chiG(\Phi_\theta,N_\theta)\,d\theta .
\label{memoryintegral}
\ee
This limiting capacity is shown in Fig.~\ref{capacity_fig}(b); for the parameters considered there, correlated noise permits a larger rate than independent noise with the same bulk single-use marginals. Appendix~\ref{memory_app} establishes that the limiting rate is operationally achievable along the growing sequence of blocks.

The same reduction applies more broadly whenever the position and momentum noise matrices commute. This includes the circulant Gauss--Markov model of Ref.~\cite{Schafer2009} and, more generally, stationary models of this form: their known finite-block Gaussian Holevo optima~\cite{Schafer2009,Schafer2011} are therefore also unrestricted optima. For a growing sequence of correlated blocks, however, identifying the finite-block optimum is not by itself sufficient to establish an operational asymptotic capacity; one must also control the limiting mode distribution and the associated coding and energy constraints. These additional conditions are treated in Appendix~\ref{memory_app}.

\section{Communication above capacity: decoding error and strong converse}\label{strong_converse_sec}

Having determined the capacity, we next examine the average outputs of optimal and near-optimal signal ensembles and the rates achievable when a nonzero decoding error is allowed. The decoding error is the receiver's probability of misidentifying the transmitted message, and we retain the same mean-input-energy constraint as in the vanishing-error problem.

\subsection{Optimal and near-optimal average outputs}

Quantum relative entropy provides a useful description of optimal signal ensembles~\cite{SchumacherWestmoreland2001}. In our setting, the capacity-achieving Gaussian average output supplies a reference state against which the average output of any ensemble can be compared. Let $H_n=\sum_{j=1}^n a_j^\dag a_j$ be the total photon-number operator, and let $D(\rho\|\sigma)\coloneqq\Tr[\rho(\ln\rho-\ln\sigma)]$ denote the quantum relative entropy.

\begin{proposition}[(Tight Gaussian reference)]\label{radius_prop}
Fix a single-mode Gaussian channel in the canonical form~\eqref{canonical} and $N>0$. Let $\sigma_*$ be the centred Gaussian average output of an optimal ensemble at energy $N$, and let $\lambda = \partial_N C(\Phi,N)$. Then, for every $n$-mode input $\rho$ with finite energy,
\bb
D\Bigl(\Phi^{\otimes n}(\rho)\Big\|\sigma_*^{\otimes n}\Bigr) \leq  nC(\Phi,N) + \lambda\big(\Tr[\rho H_n] - nN\big).
\label{radius}
\ee
In particular, the supremum of the left-hand side over $\Tr[\rho H_n] \leq  nN$ equals $nC(\Phi,N)$.
\end{proposition}

The proof in Appendix~\ref{radius_app} combines the entropy bound of Theorem~\ref{entropy_thm} with the logarithm of $\sigma_*$ to bound the relative entropy in terms of the input second moments. At the capacity-achieving point, the optimality condition makes the coefficients of the two quadrature moments coincide, so their contribution depends only on the total input photon number.

Under the assumptions of Proposition~\ref{radius_prop}, consider any ensemble $(p_i,\rho_i)_i$ with average $\bar\rho=\sum_i p_i\rho_i$ satisfying $\Tr[\bar\rho H_n]\leq nN$. Write $\bar\sigma_n=\Phi^{\otimes n}(\bar\rho)$ and $\chi_n=\chi\big(\Phi^{\otimes n};(p_i,\rho_i)_i\big)$. Averaging Eq.~\eqref{radius} and separating off the Holevo information gives
\bb
D\big(\bar\sigma_n\big\|\sigma_*^{\otimes n}\big)
\leq nC(\Phi,N)-\chi_n \eqqcolon \Delta_{\chi,n}.
\label{average_output_deficit}
\ee
The short derivation, which also applies to continuous ensembles, is given in Appendix~\ref{radius_app}. An ensemble attaining $\chi_n=nC(\Phi,N)$ must therefore have $\bar\sigma_n=\sigma_*^{\otimes n}$: the optimal average output is unique, although its ensemble decomposition need not be.

More generally, Pinsker's inequality in our natural-logarithm convention gives $\frac12\|\bar\sigma_n-\sigma_*^{\otimes n}\|_1\leq\sqrt{\Delta_{\chi,n}/2}$, so a small total Holevo deficit implies closeness of the full average output in trace distance. A vanishing per-use deficit, $\Delta_{\chi,n}/n\to0$, instead guarantees vanishing relative entropy per use, but need not imply global trace-distance convergence. The one-mode average-output marginals nevertheless converge to $\sigma_*$ on average over the channel uses; Appendix~\ref{radius_app} gives the quantitative bound.

\subsection{Exact nonzero-error capacity}

We now turn to the reliability of communication under the mean-energy constraint. Consider an $n$-use code with $M_n$ equiprobable messages, average decoding error $\epsilon_n$, success probability $p_{\rm succ}^{(n)}=1-\epsilon_n$, and rate $R_n=n^{-1}\ln M_n$; we suppress the superscript on $p_{\rm succ}$ when the block length is fixed. The input energy averaged over the messages is required to be at most $nN$. For $0\leq\epsilon<1$, let $C_\epsilon(\Phi,N)$ be the supremum of rates $R$ for which a sequence of such codes satisfies $\liminf_{n\to\infty}R_n\geq R$ and $\limsup_{n\to\infty}\epsilon_n\leq\epsilon$.

\begin{theorem}[(Exact nonzero-error classical capacity)]\label{epsilonexact_cor}
Under the mean-input-energy constraint, with both energy and decoding error averaged over equiprobable messages, every single-mode Gaussian channel $\Phi$ satisfies
\bb
C_\epsilon(\Phi,N) = C\!\left(\Phi,\frac{N}{1 - \epsilon}\right),
\qquad N\geq0,\quad 0\leq\epsilon<1.
\label{exactepsilon}
\ee
\end{theorem}

Allowing an average decoding error $\epsilon$ therefore changes the energy argument of the vanishing-error capacity from $N$ to $N/(1-\epsilon)$; the resulting rate is generally different from $C(\Phi,N)/(1-\epsilon)$. This determines the entire asymptotic rate--error trade-off under the original mean-energy constraint, including the regime in which the input energy is below $N_{\rm th}$ in~\eqref{threshold}, as shown in Fig.~\ref{converse_fig}(a).

The achievable direction follows by diluting a reliable higher-energy code with the vacuum, as in Eq.~\eqref{dilution}, while the converse proves that arbitrary quantum encodings cannot improve this trade-off. The proof in Appendix~\ref{error_app} combines that construction with the photon-number-cutoff converse of Proposition~\ref{strong_sc_thm} below and a projection estimate, Proposition~\ref{transfer_prop}. The estimate transfers the cutoff converse to mean-energy-constrained codes while retaining coherences between the low- and high-photon-number sectors.

\subsection{A complementary converse under a photon-number cutoff}

The same dilution construction explains why a mean-energy constraint does not, in general, admit a strong converse. Whenever $C(\Phi,N)<R<C(\Phi,N')$ for some $N'>N>0$, mixing each codeword of a reliable code at energy $N'$ with vacuum, with weight $N/N'$ on the original codeword, gives a code at mean energy $N$ whose success probability stays at least $N/N'$ asymptotically. Thus success need not tend to zero above $C(\Phi,N)$. This obstruction occurs even with a separate mean-energy bound for each codeword, as shown for the pure-loss channel~\cite{WildeWinter2014}.

A strict photon-number cutoff excludes those higher-energy components by requiring every codeword to be supported on the span of $n$-mode Fock states with total photon number at most $\lfloor nE\rfloor$. Under this stronger constraint the vanishing-error capacity is unchanged, but the success probability decays exponentially at every rate above it.

\begin{proposition}[(Exponential strong converse under a photon-number cutoff)]\label{strong_sc_thm}
For every single-mode Gaussian channel $\Phi$, every $E>0$, and every $R>C(\Phi,E)$, there exist $a>0$ and $n_0$ such that every $M_n$-message code with $\ln M_n \geq nR$, whose codewords are all supported on the span of $n$-mode Fock states with total photon number at most $\lfloor nE\rfloor$, satisfies
\bb
p_{\rm succ}^{(n)}<e^{-an},\qquad n \geq  n_0.
\label{strong_sc}
\ee
The encodings may be non-Gaussian and entangled across all channel uses, and the decoder may be collective. The vanishing-error capacity under this cutoff equals $C(\Phi,E)$. At $E = 0$, all inputs are vacuum and $p_{\rm succ}^{(n)} = 1/M_n$.
\end{proposition}

Proposition~\ref{strong_sc_thm} extends the known photon-number-constrained converses for phase-insensitive channels~\cite{Bardhan2015} to the exact capacity of every single-mode Gaussian channel at every input energy. Its energy-filter proof uses the capacity bound of Corollary~\ref{capacity_cor} and is given in Appendix~\ref{energyfilter_app}; cutoff achievability is established in Appendix~\ref{cutoffach_app}.

The energy-filter proof also yields a computable exponent $\mathcal{E}_{\rm filt}(R,E)>0$ for every $R>C(\Phi,E)$ under the photon-number cutoff, including when $E<N_{\rm th}$ in~\eqref{threshold}; an explicit expression is given in Appendix~\ref{filterquant_app}. It applies to arbitrary codewords satisfying the cutoff, although its constants are conservative. A separate photon-number-tail argument gives the simpler exponent $\mathcal{E}_{\rm ph}$ derived in Appendix~\ref{photon_sc_app}, which is positive whenever $R>U(\Phi,E)$, where
\bb
N_{\rm out}(E)&=kE+\frac{y(\xi+\xi^{-1})+2k-2}{4},\\
U(\Phi,E)&=G(2N_{\rm out}(E)+1)-S_{\min}.
\label{scthreshold}
\ee

\begin{figure*}[t]
\includegraphics[width=0.96\textwidth]{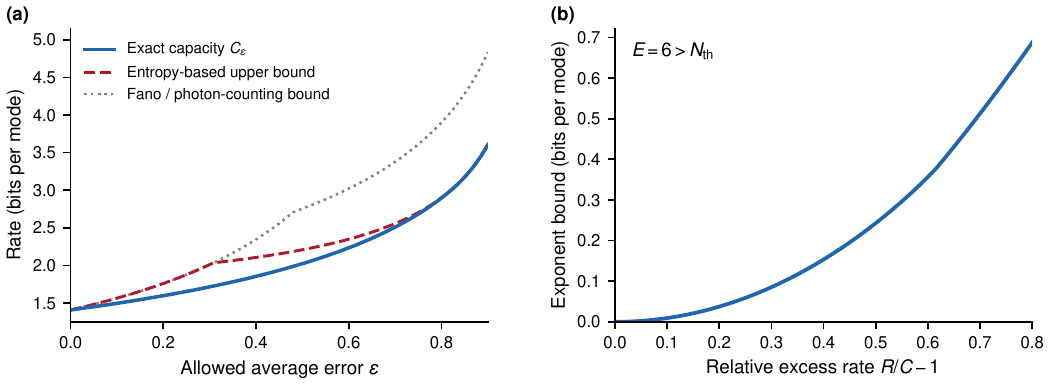}
\caption{Exact mean-energy rate--error trade-off and a complementary photon-number-cutoff converse exponent for the channel of Fig.~\ref{capacity_fig}(a). {\bf (a)} For mean input energy $N=1$, the solid blue curve is the exact nonzero-error capacity $C_\epsilon(\Phi,N)=C(\Phi,N/(1-\epsilon))$ of Theorem~\ref{epsilonexact_cor}. The dashed red curve is the weaker upper bound $\min\{C(\Phi,N)/(1-\epsilon),G(2N/(1-\epsilon)+1),U(\Phi,N/(1-\epsilon))\}$ from Fano's inequality, photon counting, and the output-entropy estimate, while the dotted grey curve is the same bound without $U$. {\bf (b)} Under the stronger requirement that every codeword be supported on the span of Fock states with total photon number at most $\lfloor6n\rfloor$, the explicit exponent bound~\eqref{photonexponent} is shown as a function of the rate relative to $C(\Phi,6)$. Here the input energy per use $E=6$ exceeds $N_{\rm th}$ in~\eqref{threshold}, so this bound is positive at every strictly positive excess rate. Proposition~\ref{strong_sc_thm} establishes exponential decay above capacity under the photon-number cutoff at every input energy, including below $N_{\rm th}$.}
\label{converse_fig}
\end{figure*}

For a Gaussian channel in the canonical form~\eqref{canonical}, $U(\Phi,E)=C(\Phi,E)$ whenever $E\geq N_{\rm th}$, with $N_{\rm th}$ given by~\eqref{threshold}. Thus, above threshold, the simpler photon-tail bound is already positive at every rate strictly above capacity under the cutoff. When $E<N_{\rm th}$, one can instead have $U(\Phi,E)>C(\Phi,E)$, so this simpler bound covers only part of the above-capacity region, whereas $\mathcal{E}_{\rm filt}$ remains positive throughout. Figure~\ref{converse_fig}(b) illustrates the photon-tail exponent. For comparison, a separate R\'enyi calculation gives an explicit exponent for the optimal Gaussian signal family when the displacement energy of each codeword is constrained (Appendix~\ref{gauss_sc_app}); these Gaussian codewords need not have strict photon-number support. None of these exponent bounds is claimed to be optimal.

For the lossy memory channel of Fig.~\ref{capacity_fig}(b), the exact mean-energy rate--error relation extends to the growing sequence of correlated blocks:
\bb
C_{{\rm mem},\epsilon}(N)
=
C_{\rm mem}\!\left(\frac{N}{1-\epsilon}\right),
\qquad N\geq0,\quad 0\leq\epsilon<1.
\label{memoryepsilon}
\ee
Here $C_{{\rm mem},\epsilon}(N)$ denotes the capacity along this sequence under the mean-energy constraint and an average decoding error at most $\epsilon$. Under a photon-number cutoff, the decoding success probability instead decays exponentially at every rate above $C_{\rm mem}(E)$. The normal-mode dilations obey bounds that are uniform at fixed memory strength and noise, allowing the energy-filter argument to pass to the limit (Appendix~\ref{memory_sc_app}). More generally, the same method gives a photon-number-cutoff converse and the corresponding mean-energy rate--error relation for any fixed finite-mode Gaussian channel in terms of its regularised Holevo capacity, even when that capacity has no known explicit formula (Appendix~\ref{filtermultimode_app}).

\section{Discussion and outlook}\label{discussion_sec}
The result identifies the optimal transmitter for single-mode Gaussian links throughout the energy range where squeezing and random displacements compete for the available energy. It also establishes the same conclusion jointly for different links sharing a power budget. For the correlated-noise channels considered here, independent squeezed signals are optimal in the normal-mode basis; transforming back can produce entanglement between physical channel uses. Thus the absence of an advantage from entangled inputs for independent links is consistent with the usefulness of collective encoding in a channel with memory.

For channels in the canonical form at positive energy, the optimal average output is also unique, and the total Holevo deficit bounds the relative entropy of any near-optimal average output from this reference. This constrains the ensemble average without fixing its decomposition into individual signals.

Our results determine the entire rate--error trade-off under the same mean-energy constraint used to define the capacity: allowing an average decoding error $\epsilon$ replaces the energy argument $N$ by $N/(1-\epsilon)$. Under a strict photon-number cutoff, the complementary strong converse forces decoding success to vanish exponentially at every rate above capacity. Both conclusions extend to the correlated-noise family studied here, including the limit of growing memory blocks, and complement recent results on energy-constrained quantum and private communication over bosonic Gaussian channels~\cite{Pirandola2026Energy,Pirandola2026Noisy,Wilde2026Quantum}.

The photon-number-cutoff strong-converse method extends beyond the channels whose capacity we can evaluate explicitly. For a fixed general multimode Gaussian channel it applies at the regularised Holevo threshold; determining that threshold remains an entropy-optimisation problem. Recent studies of coherent signals with Gaussian random displacements~\cite{PoplawskaJarzyna2026} give useful benchmarks. Within the single-mode setting, the remaining quantitative problem is to determine sharp converse exponents under the cutoff and their behaviour near capacity. Our explicit bounds establish positivity throughout the energy range, but need not describe the optimal decay rate. The R\'enyi approach of Cheng and Tomamichel~\cite{ChengTomamichel2026} gives another route to strong converses in finite dimensions; adapting it here would require energy-constrained estimates uniform in the block length as the R\'enyi order approaches one. The exact capacity and rate--error curves provide a common benchmark for assessing how efficiently optical transmitters use their available energy and how much reliability is lost when their communication rate is increased.

\begin{acknowledgments}
We acknowledge the use of AI tools: after we realised the relevance of the additivity results in Ref.~\cite[version~3]{Adesso2026} for the classical capacity problem, ChatGPT (GPT 6 Astra by OpenAI) assisted with lifting the single-use witness of Ref.~\cite{Ji2026} to many uses, leading to Proposition~\ref{epr_prop}; Claude (Opus 5.5 by Anthropic) assisted with the initial derivation of Lemma~\ref{witness_lemma} for arbitrary seeds; the present proof, however, has been obtained through heavy human intervention and independent reformulation. The same AI tools were also used to validate additional results and numerical checks and to support drafting the manuscript. We directed the research, checked every step, substantially rewrote most of the mathematical and physical content, and take full responsibility for this work.

GA acknowledges financial support from the Engineering and Physical Sciences Research Council (EPSRC Grant No.~UKRI6536). LL acknowledges financial support from the European Union under the European Research Council (ERC Grant Agreement No.~101165230).
\end{acknowledgments}

\setcounter{secnumdepth}{3}

\appendix

\section{Entropy inequality and Gaussian dilation}\label{entropy_app}
For modes $A,B$ with annihilation operators $a,b$ and positive coefficients $t,s$, the quadratic observable we need is simply
\bb
&(ta^\dag - sb)(ta - sb^\dag)\\
&\quad = \frac12\left[(tx_A - sx_B)^2 + (tp_A + sp_B)^2 + (s^2 - t^2)\id\right] \, .
\label{ordered_epr}
\ee
We parametrise a two-mode squeezed vacuum by its mean photon number $\nu\geq0$ on either mode:
\bb
|\Psi_\nu\rangle_{AB} \coloneqq \frac{1}{\sqrt{\nu + 1}}\sum_{n = 0}^{\infty}\left(\frac{\nu}{\nu + 1}\right)^{n/2}|n\rangle_A|n\rangle_B \, .
\label{epr_reference}
\ee
Its reduced states are thermal with covariance matrix $(2\nu + 1)\id$ and entropy
\bb
g(\nu) \coloneqq (\nu + 1)\ln(\nu + 1) - \nu\ln\nu = G(2\nu + 1) \, ,
\label{thermal_entropy}
\ee
where $0\ln0 = 0$. The expectation of~\eqref{ordered_epr} on $\Psi_\nu$ is $(s\sqrt{\nu + 1} - t\sqrt\nu)^2$. We use the many-pair inequality of~\cite[Theorem~2, version~3]{Adesso2026}, with a common rescaling of $t,s$ that absorbs its prefactor, and extend it to mixed states by concavity. Its one-pair case follows from~\cite[Theorem~1]{Adesso2026}, allowing an exchange of the two modes.

\begin{proposition}[(Affine entropy bound)]\label{epr_prop}
Let $A = A_1\ldots A_m$ and $B = B_1\ldots B_m$ be two collections of modes, with annihilation operators $a_j,b_j$. For each $j$, let $\nu_j,t_j,s_j>0$ satisfy
\bb
\frac{t_js_j(2\nu_j + 1)}{\sqrt{\nu_j(\nu_j + 1)}} - t_j^2 - s_j^2 = \ln(1 + 1/\nu_j) \, .
\label{epr_normalisation}
\ee
Then every state $\rho_{AB}$ with finite mean energy satisfies
\bb
S(\rho_A) \geq \sum_{j = 1}^m\bigg\{&g(\nu_j) + \left(s_j\sqrt{\nu_j + 1} - t_j\sqrt{\nu_j}\right)^2\\
&- \Tr\Big[\rho_{AB}\big(t_ja_j^\dag - s_jb_j\big)\big(t_ja_j - s_jb_j^\dag\big)\Big]\bigg\}\, .
\label{epr}
\ee
Equality holds for $\rho_{AB} = \bigotimes_j \ketbra{\Psi_{\nu_j}}$.
\end{proposition}

\begin{proof}
For pure states this is the cited inequality, after changing its squeezing parameter to $\sqrt{\nu_j/(\nu_j + 1)}$ and rescaling the two coefficients. Indeed, positivity of the left-hand side of~\eqref{epr_normalisation} is equivalent to $\sqrt{\nu_j/(\nu_j + 1)}<t_j/s_j<\sqrt{(\nu_j + 1)/\nu_j}$. Any ratio in this interval can be normalised by a common positive rescaling, and~\eqref{epr_normalisation} makes the entropy prefactor equal to one. The two possible mode orders in the original statement are included because exchanging the modes and coefficients changes~\eqref{ordered_epr} only by a constant, which cancels against its reference expectation.

For a mixed state, apply the pure-state inequality to its spectral decomposition and average, using entropy concavity and linearity of the quadratic expectations. Every eigenvector of nonzero weight has finite energy. Moreover, $(t_ja_j^\dag - s_jb_j)(t_ja_j - s_jb_j^\dag)\leq2t_j^2a_j^\dag a_j + 2s_j^2(b_j^\dag b_j + \id)$ as quadratic forms, so all expectations are finite and the countable averages are legitimate. The stated equality follows from the reference expectations above.
\end{proof}

\subsection{Choice of the environmental mode and coefficients}\label{witness_app}
We construct a dilation of $\Phi$ on the input mode $A$ and two auxiliary modes $B,E$, initially in vacuum. The auxiliary modes are chosen so that, when the input is $\gamma_u$, squeezing the receiver gives $\Psi_\nu$ on $AB$ and vacuum on $E$. We then choose $t,s$ so that, for an arbitrary input, the expectation of $(ta^\dag - sb)(ta - sb^\dag)$ depends on only one input quadrature moment. The one-pair case of Proposition~\ref{epr_prop} determines the coefficient of that dependence.

\begin{lemma}[(Quadrature-dependent expectation)]\label{witness_lemma}
Let $\Phi$ be as in~\eqref{canonical}, fix $u>0$ with $u\neq1/\xi$, and let $2\nu + 1$ be the symplectic eigenvalue of the covariance matrix of $\Phi(\gamma_u)$, so that $g(\nu) = h(u,1/u)$. There exist a symplectic unitary $V$ on three modes $ABE$ and a squeezing unitary $U_r$ on $A$ such that:
\begin{enumerate}[(a)]
\item $V$ is a Stinespring dilation of $\Phi$, meaning that
\bb
\Phi(\rho) &= \Tr_{BE}\Big[V\big(\rho_A\otimes\ketbra{0}_{BE}\big)V^\dag\Big]\, ;
\ee
and
\item defining $\widetilde V_u \coloneqq (U_r\otimes\id_{BE})V$, it holds that
\bb
\widetilde{V}_u\big(\gamma_u\otimes\ketbra{0}_{BE}\big)\widetilde V_u^\dag &= \ketbra{\Psi_\nu}_{AB}\otimes \ketbra{0}_E\, .
\label{witness_reference}
\ee
\end{enumerate}
Here $A$ labels both the input and the receiver. There also exist $t,s>0$ satisfying~\eqref{epr_normalisation} with parameter $\nu$, for which every input $\rho$ with finite second moments obeys
\bb
\Tr\!\Big[&\widetilde{V}_u \big(\rho_A\!\otimes\ketbra{0}_{BE}\big)\widetilde{V}_u^\dag\big(ta^\dag \!-\! sb\big)\big(ta \!-\! sb^\dag\big)\Big] - \left(s\sqrt{\nu \!+\!1} \!- t\sqrt\nu\right)^2\\
&= \begin{cases}
-\left.\dfrac{d}{dv}h(v,1/v)\right|_{v = u}\big(2\Tr[\rho x^2] - u\big),&u<1/\xi,\\[2mm]
-\left.\dfrac{d}{dw}h(1/w,w)\right|_{w = 1/u}\big(2\Tr[\rho p^2] - 1/u\big),&u>1/\xi.
\end{cases}
\label{pullback}
\ee
The annihilation operators $a,b$ refer to output modes $A,B$. All choices depend only on the channel and $u$, independently of $\rho$.
\end{lemma}

\begin{proof}
We order the quadratures as
\bb
R_{ABE} = (x_A,p_A,x_B,p_B,x_E,p_E)^{\mathsf T} ,
\ee
and write $\omega = \left(\begin{smallmatrix}0&1\\-1&0\end{smallmatrix}\right)$, $\Omega_3 = \omega^{\oplus3}$, and $\sigma_z = \diag(1,-1)$. The unitary $V$ corresponds to a real symplectic matrix $T$, defined by the relation $V^\dag R_{ABE}V = TR_{ABE}$. Following~\cite[Sec.~5.3]{BUCCO}, we see that, in order for $V$ to be a Stinespring dilation of $\Phi$, we may choose the first two rows, corresponding to system $A$, as follows in terms of the $X$ and $Y$ matrices of the channel:
\bb
T &= \begin{pmatrix}
X&\cos(\theta)\,\sqrt Y&-\sin(\theta)\,\sqrt Y\,\sigma_z \\
J & K & L \\
*&*&*
\end{pmatrix},
\ee
where the angle $\theta\in[0,\pi/2]$ is chosen so that\footnote{Note that the definition after Eq.~(5.48) in~\cite{BUCCO} has a typo; it should feature $\cos(2\theta_j)$ instead of $\sin(2\theta_j)$.}
\bb
\cos(2\theta) &= \frac{1-\det X}{\sqrt{\det Y}} = \frac{1 - \varsigma k}{y}\, .
\label{witness_dilation_matrix}
\ee

Complete positivity ensures that this choice of $\theta$ is possible. We are not interested, for now, in the last two ($E$) rows of $T$; however, we need to choose the blocks $J$, $K$, $L$ so that~\eqref{witness_reference} is satisfied.

Let $U_r$ be represented by the symplectic matrix
\bb
S_r \coloneqq \diag(r,r^{-1}),\qquad r \coloneqq \left(\frac{k/u + y\xi}{ku + y/\xi}\right)^{1/4} \, ,
\label{witness_squeezing}
\ee
in the sense that $U_r^\dag R_A U_r = S_r R_A$. Then, $S_r(XV_uX^{\mathsf T} + Y)S_r^{\mathsf T} = (2\nu + 1)\id_2$. Moreover, $(2\nu + 1)^2 = k^2 + y^2 + ky(\xi u + 1/(\xi u))>(k + y)^2\geq1$, so $\nu>0$. Since the symplectic matrix representing $\widetilde{V}_u$ is $(S_r\oplus\id_4)T$, we thus need to impose that
\bb
&(S_r\oplus\id_4)T(V_u\oplus\id_4)T^{\mathsf T}(S_r\oplus\id_4) \\
&\quad = \begin{pmatrix}
(2\nu + 1)\id_2&2\sqrt{\nu(\nu + 1)}\,\sigma_z&0\\
2\sqrt{\nu(\nu + 1)}\,\sigma_z&(2\nu + 1)\id_2&0\\
0&0&\id_2
\end{pmatrix}.
\label{witness_reference_covariance}
\ee
Its first two rows, multiplied on the right by $(S_r^{-1}\oplus\id_4)T^{-\mathsf T}\Omega_3$, give the identity
\bb
&\begin{pmatrix}S_r&0&0\end{pmatrix}T(V_u\oplus\id_4)\Omega_3\\
&\ = (2\nu \!+\! 1)\omega\begin{pmatrix}S_r&0&0\end{pmatrix}T + 2\sqrt{\nu(\nu \!+\! 1)}\,\sigma_z\omega\begin{pmatrix}J&K&L\end{pmatrix},
\label{witness_row_elimination}
\ee
where we used $T^{-\mathsf T} = -\Omega_3T\Omega_3$ and $S_r^{-1}\omega = \omega S_r$. Since $(\sigma_z\omega)^2 = \id_2$, this shows that
\bb
&\begin{pmatrix}J&K&L\end{pmatrix}\\
&\, = \frac{\sigma_z}{2\sqrt{\nu(\nu \!+\! 1)}}\Big[ (2\nu \!+\! 1) S_r\begin{pmatrix} X & \cos(\theta) \sqrt{Y} & -\sin(\theta) \sqrt{Y} \sigma_z\end{pmatrix} \\
&\, \quad + \omega S_r\begin{pmatrix} X & \cos(\theta) \sqrt{Y} & -\sin(\theta) \sqrt{Y} \sigma_z\end{pmatrix} (V_u\oplus\id_4)\Omega_3\Big].
\label{witness_environmental_row}
\ee

To check that this choice actually permits the desired completion, note that the first two rows of $(S_r\oplus\id_4)T$ have symplectic products $\omega$ and give covariance matrix $(2\nu + 1)\id_2$ on the input $\gamma_u\otimes\ketbra{0}_{BE}$. Together with $(V_u\oplus\id_4)\Omega_3(V_u\oplus\id_4)=\Omega_3$, these identities determine the cross terms when substituting~\eqref{witness_environmental_row}. In the symplectic product of the two-term expression in square brackets with its transpose, the two square terms give $(2\nu + 1)^2\omega$ and $\omega$, while each mixed term gives $-(2\nu + 1)^2\omega$. Hence
\bb
\begin{pmatrix}J&K&L\end{pmatrix}\Omega_3\begin{pmatrix}J&K&L\end{pmatrix}^{\mathsf T} &= \frac{1 - (2\nu + 1)^2}{4\nu(\nu + 1)}\,\sigma_z\omega\sigma_z = \omega \, .
\label{block_identities_T_1}
\ee
For the covariance matrix products, the mixed terms before multiplication by the coefficient $2\nu + 1$ are both $-\id_2$. The same substitution therefore gives the other three identities:
\begin{align}
\begin{pmatrix}S_r&0&0\end{pmatrix}T\Omega_3\begin{pmatrix}J&K&L\end{pmatrix}^{\mathsf T} &= 0, \label{block_identities_T_2}\\
\begin{pmatrix}J&K&L\end{pmatrix}(V_u\oplus\id_4)\begin{pmatrix}J&K&L\end{pmatrix}^{\mathsf T} &= (2\nu + 1)\id_2, \label{block_identities_T_3}\\
\begin{pmatrix}S_r&0&0\end{pmatrix}T(V_u\oplus\id_4)\begin{pmatrix}J&K&L\end{pmatrix}^{\mathsf T} &= 2\sqrt{\nu(\nu + 1)}\,\sigma_z \, .
\label{block_identities_T_4}
\end{align}
Due to~\eqref{block_identities_T_1}--\eqref{block_identities_T_2}, the first four rows of $T$, corresponding to modes $A$ and $B$, span a symplectic subspace. Hence, choosing a canonical pair in its symplectic complement completes $T$ to a symplectic matrix. Note that we still have some freedom in choosing $T$, as we can multiply the sub-matrix obtained by collecting the last two ($E$) rows by a local symplectic matrix on the third mode.

Due to~\eqref{block_identities_T_3}--\eqref{block_identities_T_4}, the two matrices in~\eqref{witness_reference_covariance} coincide in the $4\times 4$ top left entries, i.e.\ in the whole $AB$ block. Since this is the covariance matrix of a pure Gaussian state, $AB$ is uncorrelated with $E$. The full output is pure as well, because the input is pure and $T$ is symplectic, so the state of $E$ is also pure. By Williamson's theorem, we can transform it into the vacuum by applying a symplectic unitary on $E$ (without changing the first four rows). This establishes~\eqref{witness_reference_covariance}.

The first two columns of~\eqref{witness_environmental_row} give
\bb
J &= \frac{\sigma_z\big((2\nu + 1)S_rX + \omega S_rXV_u\omega\big)}{2\sqrt{\nu(\nu + 1)}},
\ee
Since $k>0$ and $r>0$, the two diagonal entries of $S_rX = \sqrt{k}\diag(r,\varsigma/r)$ are nonzero. We can therefore multiply by its inverse to obtain
\bb
\sigma_zJ(S_rX)^{-1} &= \frac{(2\nu + 1)\id_2 - \varsigma\diag\big((r^2u)^{-1},r^2u\big)}{2\sqrt{\nu(\nu + 1)}} \, .
\label{witness_input_block}
\ee
Since $r^4u^2 - 1 = y(\xi u^2 - \xi^{-1})/(ku + y/\xi)$, the two positive numbers $r^2u$ and $(r^2u)^{-1}$ are reciprocal and unequal. Exactly one diagonal entry of the last matrix belongs to $\left(\sqrt{\nu/(\nu + 1)},\sqrt{(\nu + 1)/\nu}\right)$: the second for $u<1/\xi$, and the first for $u>1/\xi$. Choose $t/s$ to equal that entry. Then
\bb
tS_rX - s\sigma_z J\ \propto\ \begin{cases}
\diag(1,0),&u<1/\xi,\\
\diag(0,1),&u>1/\xi,
\end{cases}
\label{witness_one_quadrature}
\ee
The proportionality factor is nonzero: the two diagonal entries in~\eqref{witness_input_block} are different, so choosing $t/s$ to equal one of them makes exactly one diagonal entry of $t\id_2 - s\sigma_zJ(S_rX)^{-1}$ vanish, and multiplying by the invertible diagonal matrix $S_rX$ preserves this property. The strict interval makes the left-hand side of~\eqref{epr_normalisation} positive before normalisation, so a common positive rescaling of $t,s$ imposes that equation.

To evaluate the quadratic observable, we write its two quadrature combinations in terms of the input quadratures. The definition of $T$ gives
\bb
\widetilde V_u^\dag\begin{pmatrix}tx_A - sx_B\\tp_A + sp_B\end{pmatrix}\widetilde V_u &= (tS_rX - s\sigma_zJ)R_A\\
&\quad + \big(t\cos(\theta)S_r\sqrt Y - s\sigma_zK\big)R_B\\
&\quad - \big(t\sin(\theta)S_r\sqrt Y\sigma_z + s\sigma_zL\big)R_E \, .
\label{witness_quadratures}
\ee
On the input $\rho_A\otimes\ketbra{0}_{BE}$, each cross term between an input quadrature on $A$ and an auxiliary quadrature on either $B$ or $E$ has zero expectation, because the vacuum first moments vanish. This remains true when $\rho$ is not centred. The terms involving only $B,E$ are independent of $\rho$, as is the scalar term in~\eqref{ordered_epr}; they therefore cancel when subtracting the expectation obtained with input $\gamma_u$. By~\eqref{witness_reference}, that expectation is $\left(s\sqrt{\nu + 1}-t\sqrt\nu\right)^2$, and~\eqref{witness_one_quadrature} now gives
\bb
\Tr\!\Big[&\widetilde{V}_u\big(\rho_A\!\otimes\ketbra{0}_{BE}\big)\widetilde{V}_u^\dag\big(ta^\dag\!-\!sb\big)\big(ta\!-\!sb^\dag\big)\Big] - \left(s\sqrt{\nu\!+\!1}\!-t\sqrt\nu\right)^2\\
&= \begin{cases}
c\big(2\Tr[\rho x^2] - u\big),&u<1/\xi,\\[2mm]
c\big(2\Tr[\rho p^2] - 1/u\big),&u>1/\xi.
\end{cases}
\label{witness_reference_subtraction}
\ee
Here $c>0$ is one quarter of the square of the nonzero entry of $tS_rX - s\sigma_zJ$, and is independent of $\rho$.

For $u<1/\xi$, keep $V,U_r,t,s$ fixed and apply the mixed-state one-pair case of Proposition~\ref{epr_prop} to the $AB$ marginal obtained from $\gamma_v$. Its receiver entropy is $h(v,1/v)$, so
\bb
h(v,1/v) \geq h(u,1/u) - c(v - u),\qquad v>0 \, ,
\label{witness_tangency}
\ee
with equality at $v = u$. Differentiability at this interior point therefore gives $c = -\left.\frac{d}{dv}h(v,1/v)\right|_{v = u}$. For $u>1/\xi$, the same argument with the input $\gamma_{1/w}$ gives the difference $c\big(2\Tr[\rho p^2] - 1/u\big)$ and identifies $c = -\left.\frac{d}{dw}h(1/w,w)\right|_{w = 1/u}$. This establishes~\eqref{pullback} and therefore completes the proof.
\end{proof}

\section{Canonical form}\label{canonical_app}
Let $\det X = \tau \neq 0$, $k = |\tau|$, and $Z_{\varsigma} = \diag(1,\varsigma)$ with $\varsigma = \sgn\tau$. The matrix $S = (X/\sqrt k)Z_{\varsigma}$ has determinant one and is therefore symplectic. Applying $S^{-1}$ at the output gives $S^{-1}X = \sqrt k Z_{\varsigma}$ and noise $Y' = S^{-1}YS^{-\mathsf{T}}>0$. Choose a rotation $O$ that diagonalises $Y'$, with its larger eigenvalue on $x$. The output rotation $O$, compensated by the input rotation $Z_{\varsigma}O^{\mathsf{T}}Z_{\varsigma}$, leaves the transfer matrix unchanged because $OZ_{\varsigma}(Z_{\varsigma}O^{\mathsf{T}}Z_{\varsigma}) = Z_{\varsigma}$. Writing the diagonal noise as $y\diag(\xi^{-1},\xi)$, with $0<\xi \leq 1$, gives Eq.~\eqref{canonical}; complete positivity requires $y^2 \geq (1 - \tau)^2$. Input rotations preserve the photon number of each mode, and output symplectic unitaries preserve all entropies. Products of single-mode channels can therefore be reduced mode by mode without changing the energy constraint or the quantities in Eq.~\eqref{waterfilling}.

\section{Gaussian Holevo information}\label{gaussian_app}
For admissible $(z,w)$ let $\hat u$ be the point of $[1/w,z]$ closest to $1/\xi$. The seed $\gamma_{\hat u}$, displaced by a Gaussian distribution of first moments with covariance matrix $[\diag(z,w) - V_{\hat u}]/2 \geq 0$, has average covariance matrix $\diag(z,w)$ and Holevo information $h(z,w) - h(\hat u,1/\hat u) = h(z,w) - \varepsilon(z,w)$, since $u\mapsto h(u,1/u)$ decreases up to $1/\xi$ and increases thereafter. By Theorem~\ref{holevo_thm}, no ensemble with these moments does better; continuous ensembles can be approximated by discrete ones~\cite{HolevoShirokov2006}. The function $h$ is jointly concave, as the composition of the increasing concave function $G$ with a geometric mean of affine functions, while $\varepsilon$ is convex by~\eqref{eps_affine}. Since the admissible region $z,w>0$, $zw\geq1$ is convex, it follows that
\[
\chiG(\Phi,N)
=
\max_{\substack{z,w>0,\ zw\geq1\\ z+w=4N+2}}
\big[h(z,w)-\varepsilon(z,w)\big]
\]
is concave in $N$.

Along the line $z + w = 4N + 2$, the arithmetic--geometric
mean inequality gives
\[
h(z,w) \leq
G\!\left(k(2N + 1) + \frac{y(\xi + \xi^{-1})}{2}\right).
\]
Equality holds at the balanced point
$kz + y/\xi = kw + y\xi$, where the output is isotropic,
provided that this point is physically admissible, i.e.,
$z,w>0$ and $zw \geq 1$. The balanced point satisfies
$z \geq 1/\xi$, and hence also $w \geq \xi$, exactly when
\[
4N + 2 \geq 2\xi^{-1}
+ \frac{y(\xi^{-1} - \xi)}{k},
\]
which is equivalent to $N \geq N_{\rm th}$, with
$N_{\rm th}$ defined in~\eqref{threshold}.
In this regime the point is physically admissible and
$\varepsilon = S_{\min}$ there. It therefore maximises
$h-\varepsilon$ and gives Eq.~\eqref{above_threshold}.

For $0<N<N_{\rm th}$, with $N_{\rm th}$ defined
in~\eqref{threshold}, the balanced point has $z<1/\xi$.
A feasible point with $z>1/\xi$ and $w \geq \xi$ cannot
maximise $h-\varepsilon$: decreasing $z$ and increasing $w$
by the same sufficiently small amount increases $h$ while
$\varepsilon$ remains equal to $S_{\min}$. If instead
$z>1/\xi$ and $w<\xi$, the same motion increases $h$ and
decreases $\varepsilon$. At a feasible boundary point
$z=1/\xi$, moving towards smaller $z$ also increases
$h-\varepsilon$: the derivative of the seed entropy
$u\mapsto h(u,1/u)$ vanishes at $u=1/\xi$, whereas $h$
increases towards the balanced point. Thus the maximiser
has $z_*<1/\xi$, so $\hat u=z_*$ and the optimal ensemble
uses the seed $\gamma_{z_*}$ displaced only in $p$.
At $N=0$, the only admissible point is $z=w=1$ and the
Holevo information vanishes.

\section{Proof of Corollary~\ref{capacity_cor}}\label{continuity_app}
\emph{Nondegenerate channels.} If all $\Phi_j$ have $\det X_j \neq 0$ and $Y_j>0$, reduce them to canonical form (Appendix~\ref{canonical_app}). The energy constraint reads $\sum_j(z_j + w_j) \leq 4E + 2m$, so Theorem~\ref{holevo_thm} gives $\chi(\Phi_1\otimes\cdots\otimes\Phi_m,E) \leq \max\sum_j[h_j(z_j,w_j) - \varepsilon_j(z_j,w_j)] \leq \max_{\sum N_j \leq  E}\sum_j\chiG(\Phi_j,N_j)$, and product Gaussian ensembles attain the right-hand side. Identity channels, with $\varepsilon = 0$, can be included. Statement (i) follows from Eq.~\eqref{waterfilling} for $m = n$ identical channels and the concavity of $\chiG(\Phi,\cdot)$; (ii) and (iii) follow in the same way, since Eq.~\eqref{waterfilling} is a supremal convolution of concave functions.

\emph{Degenerate channels.} We now remove the nondegeneracy assumptions used in the main text. When $X=0$, the output is independent of the input and the capacity therefore vanishes, whereas a rank-one $X$ remains nontrivial because one input quadrature is still transmitted. Channels with $\det X=0$, as well as those with singular $Y$, are obtained by approximation as follows. Let $\Phi = \Phi_{X,Y}$ be arbitrary. With $\tau = \det X$, choose $X_r\to X$ with $\tau_r = \det X_r \neq 0$, and $Y_r = Y + \eta_r\id$ with $\eta_r = |\tau_r - \tau|+1/r$. Then $Y_r + i(1 - \tau_r)J = [Y + i(1 - \tau)J]+[\eta_r\id + i(\tau - \tau_r)J] \geq 0$, with $J = \bigl(\begin{smallmatrix}0&1\\-1&0\end{smallmatrix}\bigr)$ the single-mode symplectic form, so $\Phi_r = \Phi_{X_r,Y_r}$ is a nondegenerate Gaussian channel. Its characteristic-function action converges pointwise to that of $\Phi$, so $\Phi_r^{\otimes n}(\rho)\to\Phi^{\otimes n}(\rho)$ in trace norm for every state $\rho$~\cite[Lemma~2]{Shirokov2018StrongStar}. For a fixed ensemble $(p_i,\rho_i)_i$ with average energy at most $E$, $\chi\big(\Phi;(p_i,\rho_i)_i\big) = \sum_ip_iD(\Phi(\rho_i)\|\Phi(\bar\rho))$, and the lower semicontinuity of the relative entropy~\cite{Wehrl1978}, with Fatou's lemma, gives $\chi\big(\Phi;(p_i,\rho_i)_i\big) \leq \liminf_r\chi\big(\Phi_r;(p_i,\rho_i)_i\big)$. The same holds for products of channels. Hence the left-hand side of Eq.~\eqref{waterfilling} is bounded by the $\liminf$ of the right-hand sides for $\Phi_{j,r}$. For the nondegenerate approximating channels, the Gaussian optimum established above allows the seed to be taken pure. Thus $\chiG$ is obtained by maximising
\bb
&G\!\left(\sqrt{\det(XVX^{\mathsf T} + Y)}\right)\\
&\qquad -G\!\left(\sqrt{\det(XV_{\rm s}X^{\mathsf T} + Y)}\right)
\label{gaussian_continuity_objective}
\ee
over pure seed covariance matrices $V_{\rm s}$ and average covariance matrices $V\geq V_{\rm s}$ with $\Tr[V]\leq4N + 2$. This set is compact: $\det V_{\rm s} = 1$ and the trace bound keep its eigenvalues bounded away from zero. The objective is continuous, including when an output state is pure, since $G$ is continuous at $1$. For the energy allocation in~\eqref{waterfilling}, use the fixed compact set of all pairs $(V_{{\rm s},j},V_j)$ with $V_j\geq V_{{\rm s},j}>0$, $\det V_{{\rm s},j} = 1$ and $\sum_j\Tr[V_j]\leq4E + 2m$. The sum of the objectives~\eqref{gaussian_continuity_objective} converges uniformly on this set as the channel matrices converge, so its maximum converges too. Gaussian ensembles attain it for the limit channels directly. This proves Eq.~\eqref{waterfilling} for all single-mode Gaussian channels, including additive noise with singular $Y$ and channels with $\det X = 0$.

\section{Memory channels}\label{memory_app}
In the ordering $(x_1,\dots,x_n,p_1,\dots,p_n)$, the $n$-use channel of Corollary~\ref{memory_cor} has $X = \sqrt k(\id_n\oplus\varsigma \id_n)$ and $Y = Y_x\oplus Y_p$. If $O$ is orthogonal with $OY_xO^{\mathsf{T}}$ and $OY_pO^{\mathsf{T}}$ diagonal, then $R = O\oplus O$ is symplectic and orthogonal, i.e., a passive linear-optical network, $RXR^{\mathsf{T}} = X$, and $RYR^{\mathsf{T}}$ is a direct sum of single-mode noise matrices. Writing $\mathcal{U}_R$ for the symplectic unitary channel implementing $R$, the original channel is $\mathcal{U}_R^\dag\circ(\bigotimes_l\Phi_l)\circ\mathcal{U}_R$. The passive unitary $\mathcal{U}_R$ preserves the total photon number, and Corollary~\ref{capacity_cor} applies to $\bigotimes_l\Phi_l$, including the regularised capacity of repeated blocks. For the model of Ref.~\cite{LPM2009}, $\Delta$ has eigenvalues $2\cos\theta_l$, which become equidistributed in $\theta\in(0,\pi)$; since $\chiG$ is continuous in the noise parameters and concave in $N$, the normalised water-filling sums converge to the integral in the main text, evaluated numerically by Lagrangian water-filling in $\theta\in(0,\pi/2)$, using the symmetry $\theta\to\pi - \theta$.

\subsection{The limit along growing blocks}

For fixed finite $s,T$ and $0<\eta<1$, the chain model has a continuous family of normal-mode channels $\Phi_\theta$, $0\leq\theta\leq\pi$, with bounded squeezing and noise. The limiting capacity $C_{\rm mem}(N)$ is defined in Eq.~\eqref{memoryintegral}. Since
$C(\Phi_\theta,N_\theta)=\chiG(\Phi_\theta,N_\theta)$
for every normal mode by Corollary~\ref{capacity_cor}, the same expression may equivalently be written in terms of the single-mode capacities $C(\Phi_\theta,N_\theta)$.

The convergence can also be seen from the dual expression
\bb
C_{\rm mem}(N)
=
\inf_{\mu>0}
\left\{
\mu N
+
\frac1\pi\int_0^\pi
\sup_{v\geq0}
\big[C(\Phi_\theta,v)-\mu v\big]\,d\theta
\right\}.
\ee
At each positive multiplier the optimising energies are uniformly bounded: the capacities grow at most logarithmically, uniformly in $\theta$. The finite-mode dual sums therefore converge to the integral by Riemann convergence; for $N>0$, their minimising multipliers stay in a compact subset of $(0,\infty)$. The zero-energy case is immediate. The finite-block converse and Fano's inequality bound every reliable sequence of codes by this limit. We spell out the additional achievability step, because regularising a fixed block is not the same as increasing the length of a correlated-noise block.

It suffices to approximate the supremum with a bounded, piecewise continuous allocation with a strict energy slack; vanishing allocations may be replaced by zero and the remaining positive allocations kept above a fixed positive number. These restrictions lose arbitrarily little rate by truncation, continuity and monotonicity. The resulting mode parameters and active energies range over compact sets. On each mode choose the optimal Gaussian seed and its Gaussian displacement ensemble. The channel in the normal-mode basis is exactly a product, although its factors differ.

For these independent signal ensembles, the information-spectrum coding theorem for classical--quantum channels~\cite{HayashiNagaoka2003} gives all rates below the normalised sum of their Holevo informations. One can verify the required concentration using the Petz R\'enyi divergence $D_a(\rho\|\sigma) = (a - 1)^{-1}\ln\Tr\big[\rho^a\sigma^{1 - a}\big]$ at orders $0<a<1$. Let $D$ denote the classical displacement label and $B$ the output mode. For their classical--quantum state $\Omega_{DB}$ and reference $\Omega_D\otimes\Omega_B$,
\bb
D_a(\Omega_{DB}\|\Omega_D\otimes\Omega_B)\longrightarrow I(D:B)\quad(a\uparrow1)
\ee
uniformly on the chosen compact parameter sets, where $I(D:B) = D(\Omega_{DB}\|\Omega_D\otimes\Omega_B)$ is the mutual information. Here is an explicit check. If $V$ is the conditional output covariance matrix and $W$ the average output covariance matrix, the Gaussian distribution of output displacements has covariance matrix $(W - V)/2$. Write $\mu_V = \sqrt{\det V}$ and $V^{[a]}$ for the covariance matrix of the normalised $a$th power of the centred Gaussian state (given explicitly in Appendix~\ref{gauss_sc_app}), and
\bb
t_a(V) = [((\mu_V + 1)/2)^a - ((\mu_V - 1)/2)^a]^{-1}.
\ee
The Gaussian trace formula and the elementary Gaussian displacement integral give
\bb
&\exp[(a - 1)D_a(\Omega_{DB}\|\Omega_D\otimes\Omega_B)]\\
&\qquad = \frac{t_a(V)t_{1 - a}(W)}{\sqrt{\det[(V^{[a]} + W^{[1 - a]} + W - V)/2]}}.
\label{memoryrenyi}
\ee
For a pure conditional state, $t_a(V) = 1$ and $V^{[a]} = V$. On the chosen active parameter sets, $W$ is faithful with $\mu_W$ uniformly above $1$; all covariance matrices are bounded. After cancelling the common divergence of $t_{1 - a}(W)$ and the determinant factor, the expression and its first derivative extend continuously to $a = 1$, including pure conditional states. Its logarithmic derivative there is $G(\mu_W) - G(\mu_V)$. Compactness gives the required uniform convergence. Additivity of $D_a$ for the independent modes and the usual binary-testing bound then place the spectral inf-divergence rate at their limiting average Holevo information. Gaussian displacement energies have uniformly bounded second moments; the strict energy slack allows the random code to be chosen to meet the average input-energy constraint as well as vanishing error. Equivalently, a continuous distribution of random displacements can first be approximated by finite constellations. Undoing the passive normal-mode transformation preserves total energy and decoding success. This proves achievability of Eq.~\eqref{memoryintegral} for this growing sequence of blocks.

This argument applies to the bounded continuous normal-mode family specified here. For a general stationary noise sequence, convergence of a spectral density alone is not a replacement for these coding and energy-control conditions.

\subsection{Uniform converse along growing memory blocks}\label{memory_sc_app}
\begin{proposition}[(Memory-channel strong converse and error trade-off)]\label{memory_sc_prop}
Fix finite memory strength $s$, thermal occupancy $T \geq 0$, and $0<\eta<1$ in the chain model of Ref.~\cite{LPM2009}. For every $E>0$ and $R>C_{\rm mem}(E)$, codes of rate at least $R$ whose inputs are supported on the subspace spanned by Fock states with total photon number at most $\lfloor nE\rfloor$ have exponentially vanishing success along the growing sequence of blocks. The cutoff capacity equals $C_{\rm mem}(E)$. Under a mean-energy constraint and average error $0 \leq \epsilon<1$, Eq.~\eqref{memoryepsilon} holds.
\end{proposition}
\begin{proof}
The passive normal-mode transformation preserves total photon number. It reduces each block to $\bigotimes_{j = 1}^n\Phi_{\theta_j}$, with $\theta_j = \pi j/(n + 1)$. These channels need not be identical. Each has a dilation on three modes: prepare a two-mode squeezed vacuum purifying the thermal environment, squeeze the environmental input by $s\cos\theta$, and couple it to the signal with a beam splitter of transmissivity $\eta$. If $r_T = \tfrac12\operatorname{arcosh}(2T + 1)$, the dilation matrices satisfy
\bb
\sup_{\theta\in[0,\pi]}\max\{\|S_\theta\|_\infty,\|S_\theta^{-1}\|_\infty\} \leq  e^{|s|+r_T}.
\label{memorydilation}
\ee
This remains valid at $T = 0$, with an unused vacuum purification mode.

All estimates in Appendix~\ref{energyfilter_app} are consequently uniform in $n$ and in the normal-mode parameters. To see this explicitly, the Husimi moment argument uses only a common upper bound on the operator norms of the block-diagonal symplectic matrices and their inverses. A common smoothing covariance matrix and the radial input identity therefore give a single function $\ell(s',L)$ for the whole product, allowing arbitrary correlated inputs. Here $s'>0$ is the moment parameter, distinct from the memory strength. The small-subset quadratic estimate uses the same uniform coefficient operator norm. The image projectors and transformed input Hamiltonians still commute site by site, and the polynomial degree, subset counting, and receiver Fock dimensions are unchanged. Thus the approximation error and projective-norm bounds have constants independent of the block length.

Let
\bb
C_n(E_1) = \frac1n\max_{\sum_jN_j \leq  nE_1} \sum_{j = 1}^n C(\Phi_{\theta_j},N_j).
\ee
Corollary~\ref{capacity_cor} bounds the Holevo information of every auxiliary ensemble selected by the positive filter by $nC_n(E_1)$. As proved above, $C_n(E_1)\to C_{\rm mem}(E_1)$. The limiting function is finite, concave, and continuous for $E_1>0$. Choose $E_1>E$ with $C_{\rm mem}(E_1)<R$ and apply the same filter inequality and contradiction as in Proposition~\ref{strong_sc_thm}, replacing $C(\Phi,E_1)$ by $C_n(E_1)$. Its convergence and the uniform approximation constants give one positive exponential decay rate for all sufficiently large blocks. This establishes a converse for growing blocks, without replacing them by repetitions of a fixed block.

For cutoff achievability, take a bounded normal-mode allocation with strict energy slack, as in the preceding achievability proof. The average input is a product of centred Gaussian states whose covariance matrices range over a compact set. Their photon-number generating functions are finite on a common neighbourhood of zero, with uniformly bounded second derivatives of their logarithms. Their first derivatives are their mean photon numbers. A Chernoff bound therefore makes the average weight above $nE$ exponentially small. Random coding can simultaneously make this weight and the decoding error vanish. Projecting each codeword onto the cutoff and replacing its discarded weight by vacuum changes average success by a vanishing amount. Letting the energy slack tend to zero proves cutoff achievability at $C_{\rm mem}(E)$.

Finally, Proposition~\ref{transfer_prop} applies to an arbitrary block channel, so the uniform cutoff converse gives the upper bound in Eq.~\eqref{memoryepsilon}. Diluting the reliable codes at energy $N'<N/(1 - \epsilon)$ with vacuum gives the lower bound. Continuity permits $N'\uparrow N/(1 - \epsilon)$. At $N = 0$ all inputs are vacuum and both sides vanish; at $\epsilon = 0$ the preceding capacity result applies.
\end{proof}

Uniform dilation bounds and convergence to a continuous limiting capacity are the additional ingredients beyond a fixed-block converse. The proof also covers other normal-mode arrays satisfying these conditions and the corresponding coding assumptions. It does not claim a result for arbitrary stationary noise solely from the existence of a limiting noise spectrum.

\section{The optimal Gaussian reference state}\label{radius_app}
Let $(z_*,w_*)$ maximise $h(z,w) - \varepsilon(z,w)$ under $z + w = 4N + 2$. For a nondegenerate Gaussian channel in canonical form~\eqref{canonical} and $N>0$, the optimum has a nontrivial distribution of random displacements, hence $z_*w_*>1$. The function $\varepsilon$ is differentiable there: in the two constrained regimes it is the appropriate decreasing branch of the seed entropy, while its derivatives vanish where the seed with covariance matrix $\diag(1/\xi,\xi)$ is feasible. Denote its gradient by $b = (b_x,b_p)$ at the optimum. The energy multiplier satisfies
\bb
h_z(z_*,w_*) - b_x = h_w(z_*,w_*) - b_p = \lambda/4.
\label{kkt}
\ee
The same multiplier gives the derivative of the capacity: the objective has a unique maximising covariance matrix because $h$ is strictly concave on each fixed-energy line and $\varepsilon$ is convex. The usual envelope argument applies, including the input energy $N=N_{\rm th}$ in~\eqref{threshold}, where $b\to0$.

For a state $\rho$ with finite second moments, its Gaussification $\rho_{\mathrm{G}}$ is the unique Gaussian state with the same first moments and covariance matrix. We use the following observation.

\begin{fact}[(Gaussification and Gaussian channels)]\label{gaussification_fact}
For every Gaussian channel $\Phi$ between finitely many modes and every input state $\rho$ with finite second moments,
\bb
\big(\Phi(\rho)\big)_{\mathrm{G}} = \Phi(\rho_{\mathrm{G}}) \, .
\label{gaussification}
\ee
\end{fact}
\begin{proof}
The affine transformation of first moments and covariance matrices under $\Phi$ is the same for $\rho$ and $\rho_{\mathrm{G}}$. Since $\Phi(\rho_{\mathrm{G}})$ is Gaussian, uniqueness of the Gaussian state with the resulting moments proves~\eqref{gaussification}.
\end{proof}

By Fact~\ref{gaussification_fact}, $[\Phi^{\otimes n}(\rho)]_{\mathrm{G}} = \Phi^{\otimes n}(\rho_{\mathrm{G}})$ for every input $\rho$ with finite second moments. Thus replacing the input by the Gaussian state with the same first and second moments also preserves those moments at the output. Since the logarithm of the faithful, centred Gaussian state $\sigma_*$ is a constant plus a diagonal quadratic form, this replacement preserves the cross entropy against $\sigma_*^{\otimes n}$. Equivalently, the tangent plane of Gaussian entropy gives the cross entropy exactly:
\bb
-\Tr\big[\Phi^{\otimes n}(\rho)\ln\sigma_*^{\otimes n}\big] & = nh(z_*,w_*) + h_z\sum_j(z_j - z_*)\\
&\quad + h_w\sum_j(w_j - w_*),
\label{crossentropy}
\ee
where $z_j = 2\langle x_j^2\rangle_\rho$, $w_j = 2\langle p_j^2\rangle_\rho$, and $h_z = \partial_z h$, $h_w = \partial_w h$ are evaluated at $(z_*,w_*)$. Convexity and Theorem~\ref{entropy_thm} imply
\bb
S\Big(\Phi^{\otimes n}(\rho)\Big) \geq  n\varepsilon(z_*,w_*) + b_x\sum_j(z_j - z_*) + b_p\sum_j(w_j - w_*).
\ee
Subtract this from Eq.~\eqref{crossentropy} and use Eq.~\eqref{kkt} to obtain Eq.~\eqref{radius}. It is saturated by products of optimal seeds displaced only in directions in which the random displacements have nonzero variance, with total displacement energy $n(N - N_s)$, where $N_s = (u_*+u_*^{-1} - 2)/4$ is the energy of the optimal seed. For input energy $N<N_{\rm th}$, with $N_{\rm th}$ defined in~\eqref{threshold}, the active direction is $p$; for input energy $N>N_{\rm th}$ both directions are active. This also proves the asserted supremum.

\emph{Optimal and near-optimal average outputs.} Under the assumptions of Proposition~\ref{radius_prop}, let $(p_i,\rho_i)_i$, $\bar\rho$, $\bar\sigma_n$, $\chi_n$, and $\Delta_{\chi,n}$ be as in Eq.~\eqref{average_output_deficit}. Expanding the relative entropies gives the exact identity
\[
\begin{aligned}
&\sum_i p_i D\big(\Phi^{\otimes n}(\rho_i)\big\|\sigma_*^{\otimes n}\big)\\
&\qquad = \chi_n+D\big(\bar\sigma_n\big\|\sigma_*^{\otimes n}\big).
\end{aligned}
\]
Averaging Eq.~\eqref{radius} bounds its left-hand side by
\[
nC(\Phi,N)+\lambda\big(\Tr[\bar\rho H_n]-nN\big)\leq nC(\Phi,N),
\]
because $\lambda\geq0$ and $\Tr[\bar\rho H_n]\leq nN$. This proves Eq.~\eqref{average_output_deficit}. Finite mean energy makes the entropy and quadratic cross-entropy terms finite, and the same argument applies to continuous ensembles with integrals in place of sums.

For completeness, let $\bar\sigma_{n,j}$ be the $j^{\text{th}}$ one-mode marginal of $\bar\sigma_n$. Subadditivity gives
\[
\sum_{j=1}^n D(\bar\sigma_{n,j}\|\sigma_*)
\leq D\big(\bar\sigma_n\big\|\sigma_*^{\otimes n}\big)
\leq\Delta_{\chi,n}.
\]
Pinsker's inequality and Cauchy--Schwarz therefore imply
\[
\frac1n\sum_{j=1}^n\|\bar\sigma_{n,j}-\sigma_*\|_1
\leq\sqrt{\frac{2\Delta_{\chi,n}}{n}}.
\]
Thus a vanishing per-use Holevo deficit controls the one-mode average-output marginals on average over the channel uses. It need not control every selected marginal or the trace distance of the full $n$-mode state. If the total deficit tends to zero, Pinsker's inequality instead gives global convergence through $\frac12\|\bar\sigma_n-\sigma_*^{\otimes n}\|_1\leq\sqrt{\Delta_{\chi,n}/2}$.

\section{Mean energy and nonzero error}\label{error_app}
The input energy is averaged over equiprobable messages, and the error is the average decoding error. For comparison with the exact result~\eqref{exactepsilon}, elementary estimates give, for $0<\epsilon<1$,
\bb
C\!\left(\Phi,\frac{N}{1 - \epsilon}\right) \leq  C_\epsilon(\Phi,N) \leq \min\!\left\{\frac{C(\Phi,N)}{1 - \epsilon},\,g\!\left(\frac{N}{1 - \epsilon}\right)\right\},
\label{epsiloncapacity}
\ee
Measurement data processing and the capacity theorem bound the message--decoder mutual information by $nC(\Phi,N)$. Writing $h_2(t) \coloneqq -t\ln t - (1 - t)\ln(1 - t)$, with $0\ln0 = 0$, Fano's inequality gives
\bb
(1 - \epsilon_n)R_n \leq  C(\Phi,N) + \frac{h_2(\epsilon_n)}n,
\label{fano}
\ee
and hence the first upper bound. Photon-number subspace dimension gives the second, as shown below. These comparisons are generally weaker than the exact result.

For the lower bound at $N>0$, choose any $N'<N/(1 - \epsilon)$ with $N'>N$, and a reliable code at any rate $R<C(\Phi,N')$, with codewords $\rho_m^{(n)}$ and average energy at most $nN'$. Replace each codeword by
\bb
\widetilde\rho_m^{(n)} = \frac{N}{N'}\rho_m^{(n)} + \left(1 - \frac{N}{N'}\right)|0\rangle\langle0|^{\otimes n}.
\label{dilution}
\ee
The average energy is at most $nN$. With the original decoder its success probability is at least $(N/N')p_{\rm succ}^{(n)}$, so its limiting error is strictly below $\epsilon$. Let $N'\uparrow N/(1 - \epsilon)$ and use continuity of the capacity. The argument extends the known pure-loss construction~\cite{WildeWinter2014}; the channel-specific input here is the exact phase-sensitive capacity. Equation~\eqref{dilution} remains possible under a per-message bound on the \emph{expected} photon number if the original code meets such a bound. A photon-number support condition is substantially stronger.

\subsection{A sharp projection bound}

\begin{proposition}[(Transferring photon-number bounds to mean energy)]\label{transfer_prop}
For a fixed channel $\Phi$, let $B_n(M,E)$ be the supremum of the decoding success probability for $M$-message codes whose inputs are supported on the span of $n$-mode Fock states with total photon number at most $\lfloor nE\rfloor$. Every code with average energy at most $nN$ satisfies
\bb
p_{\rm succ} \leq \min\!\left\{1, \left(\sqrt{B_n(M,E)} + \sqrt{\delta_n}\right)^2\right\},
\label{transfer}
\ee
where $\delta_n = \min\{1,nN/(\lfloor nE\rfloor + 1)\}$.
\end{proposition}

\begin{proof}
We prove the bound without dephasing the codewords. Put $\Pi = \id_{H_n \leq \lfloor nE\rfloor}$ and $Q = \id - \Pi$, and let $\{\Lambda_m\}_{m = 1}^M$ be the decoder POVM. Apply the adjoint channel to its effects, defining $T_m = (\Phi^{\otimes n})^\dag(\Lambda_m)$. Then $0 \leq  T_m \leq \id$ and $\sum_mT_m = \id$. Define
\bb
p_L = \frac1M\sum_m\Tr\big[T_m\Pi\rho_m\Pi\big],\qquad p_H = \frac1M\sum_m\Tr\big[T_m Q\rho_m Q\big].
\ee
Cauchy--Schwarz for the Hilbert--Schmidt inner product, followed by Cauchy--Schwarz over messages, bounds the cross term by $\sqrt{p_Lp_H}$. Hence
\bb
p_{\rm succ} \leq  p_L + p_H + 2\sqrt{p_Lp_H} = \left(\sqrt{p_L} + \sqrt{p_H}\right)^2.
\label{sectorbound}
\ee
In particular, $p_H \leq \Tr\big[Q\bar\rho\big] \leq \delta_n$ by the energy constraint. Replace each input by the normalised state
\bb
\rho'_m = \Pi\rho_m\Pi + \Tr[Q\rho_m]|0\rangle\langle0|^{\otimes n}.
\ee
These form a valid code supported on $\operatorname{ran}\Pi$. Its success with the same decoder is at least $p_L$, so $p_L \leq  B_n(M,E)$. Equation~\eqref{transfer} follows. The argument allows arbitrary entangled inputs and coherences between the two sectors.
\end{proof}

There is also an unconditional application to every channel. The input subspace has dimension $d_n(E) = \binom{n + \lfloor nE\rfloor}{n}$. For a code supported on $\operatorname{ran}\Pi$, $\rho_m \leq \Pi$ gives
\bb
p_{\rm succ}&\leq\frac1M\sum_m\Tr\big[T_m\Pi\big]=\frac{d_n(E)}M,\\
\lim_{n\to\infty}\frac1n\ln d_n(E)&=g(E) \, .
\ee
Using Eq.~\eqref{transfer} at every $E>N/(1 - \epsilon)$ yields $C_\epsilon(\Phi,N) \leq  g(N/(1 - \epsilon))$, the second upper bound in Eq.~\eqref{epsiloncapacity}.

The transfer also retains finite-block information. If, at a fixed rate, $B_n(M_n,E) \leq  e^{-n\kappa}$ with $\kappa>0$, then
\bb
p_{\rm succ} \leq \delta_n + 2\sqrt{\delta_n}\,e^{-n\kappa/2} + e^{-n\kappa}.
\ee
Thus the upper bound approaches the mean-energy ceiling, with an exponentially decaying remainder. This distinguishes an exponential remainder from an exponential strong converse under a mean-energy constraint.

\subsection{Proof of Theorem~\ref{epsilonexact_cor}}
The proof combines the cutoff converse of Proposition~\ref{strong_sc_thm}, the projection estimate of Proposition~\ref{transfer_prop}, and the vacuum-dilution construction~\eqref{dilution}.

For $N>0$, suppose $R>C(\Phi,N/(1 - \epsilon))$. By continuity in the energy, choose $E>N/(1 - \epsilon)$ close enough that $R>C(\Phi,E)$. Proposition~\ref{strong_sc_thm} gives $B_n(M_n,E)\to0$ for codes of asymptotic rate at least $R$, and Proposition~\ref{transfer_prop} therefore gives $\limsup_n p_{\rm succ}^{(n)} \leq N/E<1 - \epsilon$. Such codes cannot have asymptotic average error at most $\epsilon$, proving the upper bound. For $0<\epsilon<1$, vacuum dilution as in~\eqref{dilution} gives the reverse inequality; for $\epsilon = 0$, it is the definition of the vanishing-error capacity. At $N=0$, only the vacuum input is allowed and the capacity is zero. This proves~\eqref{exactepsilon}.

\section{Energy-filter proof of the photon-number-cutoff strong converse}\label{energyfilter_app}
The proof adapts the finite-dimensional method of Beigi and Tomamichel~\cite{BeigiTomamichel2026} to the input-energy constraint. In a Gaussian dilation, we construct a positive filter that fixes the images of inputs from the span of Fock states with at most $\lfloor nE\rfloor$ photons. Its range contains only images of inputs from the analogous subspace with at most $\lfloor nE_1\rfloor$ photons, where $E_1>E$. This ensures that the auxiliary codewords used in the converse obey the slightly larger energy constraint. Polynomial approximations, together with Gaussian photon-number estimates and counts of Fock states on the modes affected by each term, control both the approximation error and its projective tensor norm across receiver and environment. The resulting coding inequality rules out success decaying too slowly at rates above $C(\Phi,E_1)$; continuity then lets $E_1$ approach $E$.

The following proof keeps the input Hamiltonian throughout. By Corollary~\ref{capacity_cor}, every $n$-use ensemble of average energy at most $nE$ satisfies
\bb
\chi\big(\Phi^{\otimes n};(q_x,\rho_x)_x\big) \leq  n C(\Phi,E),
\label{sc_holevo}
\ee
where $C(\Phi,E)$ is finite and continuous for $E>0$. This is the only capacity bound used below. Set $P_{n,E} = \id_{\{H_n \leq \lfloor nE\rfloor\}}$.

After removing its output displacement, every single-mode Gaussian channel, including the degenerate canonical classes, admits a finite symplectic unitary dilation with a pure Gaussian environment~\cite{Caruso2011}. Absorbing the preparation of that pure environment into the unitary makes its auxiliary inputs vacuum without changing the input Hamiltonian.

Write the resulting symplectic unitary as $U$ on $d = 1 + r$ modes, with $r$ vacuum auxiliary modes and outputs $B,F$. We write $H_{A_j} = a_j^\dag a_j$ and $\lvert0\rangle_F$ for the auxiliary vacuum. The letter $F$ denotes the entire environment, including purification modes. Define the isometry $V\scket\psi = U(\scket\psi\otimes\scket{0}^{\otimes r})$. On the full output Fock space put
\bb
\Pi_j = U_j(\id_A\otimes\scket0\scbra0_F)U_j^\dag, \qquad K_j = U_j(H_{A_j}\otimes \id_F)U_j^\dag.
\ee
The operators $K_j$ are nonnegative quadratic bosonic operators. All the $K_j$ and $\Pi_j$ commute, including at the same site. Set $\Pi^{(n)} = \prod_j\Pi_j$ and $K^{(n)} = \sum_jK_j$.

\subsection{A positive-filter coding inequality}
For an operator $A$ on $B:F$, use the projective norm
\bb
\scnorm A_\pi = \inf\left\{\sum_l\scnorm{X_l}_\infty\scnorm{Y_l}_\infty: A = \sum_lX_l\otimes Y_l\right\}.
\ee
All approximants constructed below have finite local support, so finite sums suffice. Conjugation by local isometries does not increase this norm. Finite local support means $A = (P_B\otimes P_F)A(P_B\otimes P_F)$ for finite-rank local projections. Operator norms are denoted by $\|\cdot\|_\infty$; vector norms are Hilbert-space norms.

\begin{lemma}[(Energy-preserving positive filter)]\label{sc_filter_lemma}
Fix $E<E_1$. Let $0 \le  G \le \id$ be a finite-rank operator on the joint Stinespring output such that
\begin{enumerate}
\item $G V^{\otimes n}P_{n,E} = V^{\otimes n}P_{n,E}$;
\item $\operatorname{ran}G\subseteq V^{\otimes n}\operatorname{ran}P_{n,E_1}$.
\end{enumerate}
Suppose $A$ has finite local support, $\scnorm{A - G}_\infty \le \epsilon$, and $\scnorm A_\pi \le \Gamma$. For an $M$-message photon-number-cutoff code of success $p>0$, put $z = \epsilon/\sqrt p$. If $z<1/4$, then
\bb
n C(\Phi,E_1) \ge {}&(1 - 4z)\ln M + 2\ln p - 2\ln\Gamma + 2\ln(1 - z)\\
&-(1 + 2z)h_2\left(\frac{2z}{1 + 2z}\right).
\label{sc_filtercoding}
\ee
\end{lemma}
\begin{proof}
A receiver-side Naimark isometry makes the decoder effects orthogonal projections $Q_m$. Conjugate $G$ and $A$, and compose the channel isometry, with that local isometry; their bounds and input-energy interpretation are unchanged. Let $\lambda_m = \scnorm{GQ_mG}_\infty$. The identity action of $G$ on the original input image implies that $\lambda_m$ is at least the success of the original $m$th codeword, including for mixed codewords. For $\lambda_m>0$, choose a normalised top eigenvector $\scket{\psi_m}$ of $GQ_mG$. It belongs to $\operatorname{ran}G$ and therefore corresponds to a genuine input supported on $\operatorname{ran}P_{n,E_1}$. Also $\lambda_m \le 1$. Set $Z = \sum_m\lambda_m \ge  Mp$ and $q_m = \lambda_m/Z$; zero terms are omitted. Writing $H(X)_q = -\sum_mq_m\ln q_m$ for the label entropy, we have $H(X)_q \ge - \ln\max_mq_m \ge \ln M + \ln p$.

On two classical-label copies and the dilated output define
\bb
\scket\psi& = \sum_m\sqrt{q_m}\scket m_X\scket m_{X'}\scket{\psi_m},\\
\scket\theta& = \sum_m\frac{\sqrt{q_m}}{\lambda_m} \scket m_X\scket m_{X'}Q_mG\scket{\psi_m}.
\ee
The eigenvector equations give $G\scket\theta = \scket\psi$ and $\scnorm\theta^2 = \sum_mq_m/\lambda_m \le 1/p$. Consequently $\scnorm{A\scket\theta - \scket\psi} \le  z$ and $v \coloneqq \scnorm{A\scket\theta} \ge 1 - z$. The normalised vector $\scket\varphi = A\scket\theta/v$ has trace distance (with the factor $1/2$) at most $2z$ from $\scket\psi$.

Orthogonality of the decoder projectors implies
\bb
\theta_{XX'F} \le \id_X\otimes\theta_{X'F}, \qquad\Tr\big[\theta_{X'F}\big] = \scnorm\theta^2 \le 1/p.
\ee
For a decomposition $A = \sum_lc_lX_l\otimes Y_l$ with local contractions, $c_l \ge 0$, and $\sum_lc_l \le \Gamma$ (or arbitrarily close), apply vector Cauchy--Schwarz. For each contraction $X_l$ and positive $T$, one has $\Tr_B\big[(X_l\otimes \id)T(X_l^\dag\otimes \id)\big] \leq \Tr_B[T]$. Together these give
\bb
\varphi_{XX'F} \le \id_X\otimes\mu_{X'F},\qquad \Tr\big[\mu_{X'F}\big] \le \frac{\Gamma^2}{p v^2}.
\ee
More explicitly, one may take $\mu = v^{-2}\Gamma\sum_lc_l(\id_{X'}\otimes Y_l) \theta_{X'F}(\id_{X'}\otimes Y_l^\dag)$. Operator domination bounds the max-relative entropy, and hence the relative entropy, against $\id_X\otimes\mu/\Tr\big[\mu\big]$. Positivity of the marginal relative entropy then gives $S(X|X'F)_\varphi \ge - \ln\Tr\big[\mu\big]$. With $S(A|B) = S(AB) - S(B)$, pure-state conditional-entropy duality therefore yields
\bb
S(X|B)_\varphi \le 2\ln\Gamma - \ln p - 2\ln v.
\ee
The conditional-entropy continuity bound with finite $X$~\cite{Winter2016} gives
\bb
S(X|B)_\psi \le  S(X|B)_\varphi + 4z\ln M + (1 + 2z)h_2\left(\frac{2z}{1 + 2z}\right).
\ee
The $XB$ marginal of $\psi$ is a genuine channel ensemble with all inputs supported on $\operatorname{ran}P_{n,E_1}$. Subtract its conditional entropy from $H(X)_q$ and use \eqref{sc_holevo}, $v \ge 1 - z$, to obtain \eqref{sc_filtercoding}. The approximant has finite receiver and environment support. Conditional entropy continuity depends only on $\dim X = M$; the genuine channel ensemble also has finite energy. No finite-dimensional assumption on the full bosonic output is needed.
\end{proof}

\subsection{Uniform exponential moments for a symplectic unitary}
Let $H_{dn}^{\rm phys}$ be the total physical photon number on $dn$ modes.

\begin{lemma}\label{sc_moment_lemma}
For a fixed $d$-mode symplectic unitary $W$, there is $s_0>0$ depending only on $W$, and a finite function $\ell_W(s,L)$ for every $L \ge 0$ and $0<s<s_0$, such that
\bb
\sup_{\rho:\,\operatorname{supp}\rho\subseteq\operatorname{ran}\id_{\{H_{dn}^{\rm phys}\leq nL\}}} \Tr\big[W^{\otimes n}\rho W^{\dag\otimes n}e^{sH_{dn}^{\rm phys}}\big] \le  e^{n\ell_W(s,L)}.
\label{sc_moment}
\ee
The function is independent of $n$ and of correlations in $\rho$.
\end{lemma}
\begin{proof}
Let $S$ be the symplectic matrix of $W$, $\kappa_S = \scnorm S_\infty^2$, and choose $c \ge (\kappa_S - 1)/2$. Put $A = c\id + (\id - SS^{\mathsf{T}})/2 \ge 0$ and $a_S = \scnorm A_\infty$. Let $Q_\rho$ denote the heterodyne (Husimi) probability distribution of $\rho$. Its real outcome vector $Z$ is normalised so that the vacuum has covariance matrix $\id$, and
\bb
\mathbb{E}_{Q_\rho}e^{u|Z|^2/2} = \Tr\big[\rho(1 - u)^{-(H_{dn}^{\rm phys} + dn)}\big],\qquad 0<u<1.
\ee
Adding independent centred Gaussian noise with covariance matrix $c\id$ to the output Husimi distribution gives the distribution of $S^{\oplus n}Z + \zeta$, where $\zeta$ is independent Gaussian with covariance matrix $A^{\oplus n}$. This characteristic-function identity is valid for arbitrary, possibly non-Gaussian, inputs. Set
\bb
u = 1 - e^{-s},\qquad v = \frac{u}{1 + cu},\qquad t = \frac{v\kappa_S}{1 - va_S}.
\ee
For sufficiently small $s>0$, $va_S<1$ and $t<1$. Conditional Gaussian integration and $S^{\mathsf{T}}(\id - vA)^{-1}S \le \kappa_S(1 - va_S)^{-1}\id$ give
\bb
\mathbb{E} e^{v|S^{\oplus n}Z + \zeta|^2/2} \leq (1 - va_S)^{-dn}\mathbb{E}_{Q_\rho}e^{t|Z|^2/2}.
\ee
Undoing the classical smoothing contributes $(1 - cv)^{dn}$, while converting the output Husimi moment into the photon moment contributes $e^{-dns}$. The resulting bound is
\bb
\ell_W(s,L)& = -ds + d\ln(1 - cv) - d\ln(1 - va_S)\\
&\quad - (L + d)\ln(1 - t).
\label{sc_ell}
\ee
The remaining input Husimi moment is at most $(1 - t)^{-n(L + d)}$ by the radial identity and the input cutoff. All uses of positivity concern Husimi distributions, not Wigner functions. The nonnegative integrals may initially be read in the extended sense; the finite input bound proves their integrability.
\end{proof}

\subsection{Approximating an energy filter}
\begin{proposition}\label{sc_approx_prop}
Fix $0<E<E_1<\infty$. For every $\eta>0$ there exist $\kappa>0$, finite-rank positive contractions $G_n$, and finite-local-support operators $A_n$ such that
\bb
G_nV^{\otimes n}P_{n,E}& = V^{\otimes n}P_{n,E},\\
\operatorname{ran}G_n&\subseteq V^{\otimes n}\operatorname{ran}P_{n,E_1},\\
\scnorm{A_n - G_n}_\infty& \le  e^{-\kappa n},\qquad \scnorm{A_n}_\pi \le  e^{\eta n}
\label{sc_approx}
\ee
for all sufficiently large $n$.
\end{proposition}

\begin{proof}
\emph{Choice of output and spectral cutoffs.} Choose a sufficiently small positive moment parameter $s$ for both $U$ and $U^\dag$. By Lemma~\ref{sc_moment_lemma}, choose $F_0$ so large that $sF_0>\ell_U(s,E_1)$. Let
\bb
P = P_{B, \le \lfloor nF_0\rfloor}\otimes P_{F, \le \lfloor nF_0\rfloor}.
\ee
Its complement is contained in $\{H_{dn}^{\rm phys}>nF_0\}$. On its range, $H_{dn}^{\rm phys} \le 2nF_0$. Because $K^{(n)} \le  U^{\otimes n}H_{dn}^{\rm phys}U^{\dag\otimes n}$ and the two operators commute, Lemma~\ref{sc_moment_lemma} also implies
\bb
\scnorm{P e^{sK^{(n)}}P}_\infty \le  e^{n\ell_{U^\dag}(s,2F_0)}.
\label{sc_reverse}
\ee
Fix $T>\max\{1,E_1\}$ large enough that $sT>\ell_{U^\dag}(s,2F_0) + 2$. Here $P_{B, \le  L}$ and $P_{F, \le  L}$ project onto at most $L$ total photons in the receiver and environment, respectively. These cutoffs are fixed before choosing polynomial degrees.

\emph{Energy polynomial and exact filter.} Choose small constants $t_0,t_1>0$, put $D_i = \lfloor t_i n\rfloor$, and set $b = (E + E_1)/(2T)$. Define
\bb
q_n(x) = \sum_{j = 0}^{\lfloor D_1b\rfloor} \binom{D_1}{j}(x/T)^j(1 - x/T)^{D_1 - j}.
\label{sc_binomial}
\ee
On $[0,T]$ it is a binomial lower-tail probability and lies in $[0,1]$. Hoeffding's inequality yields, conservatively for large $n$,
\bb
|q_n(x) - 1|& \le \epsilon_1\quad(0 \le  x \le  E),\\
|q_n(x)|& \le \epsilon_1\quad(E_1 \le  x \le  T),\\
\epsilon_1& = e^{-D_1(E_1 - E)^2/(8T^2)}.
\ee
Let $f_n(x)$ equal $1$ for $0 \le  x \le  E$, equal $q_n(x)$ for $E<x<E_1$, and equal $0$ for $x \ge  E_1$. Define
\bb
G_n = \Pi^{(n)}f_n(K^{(n)}/n).
\ee
This is a finite-rank positive contraction with the exact identity and energy-support properties in the proposition. It is not required to be a projection or a Gaussian operator. Since the range of $G_n$ is contained in the image of $P_{n,E_1}$, the direct moment bound gives $\scnorm{(\id - P)G_n}_\infty^2 \leq  e^{-n[sF_0 - \ell_U(s,E_1)]}$. Using self-adjointness and $\scnorm{G_n - PG_nP}_\infty \leq 2\scnorm{(\id - P)G_n}_\infty$ yields
\bb
\scnorm{G_n - PG_nP}_\infty \le 2e^{-n[sF_0 - \ell_U(s,E_1)]/2}.
\label{sc_directtail}
\ee

\emph{Image polynomial.} Write $\Delta_j = \id - \Pi_j$ and $L_n = \sum_j\Delta_j$. A NOR polynomial approximates one when all $\Delta_j$ vanish and zero otherwise. The approximation of~\cite[Theorem~4.5]{Sherstov2020}, used in~\cite[Lemma~3 and Fact~2]{BeigiTomamichel2026}, applies to commuting projections on arbitrary Hilbert spaces: it gives a real polynomial $p_n$ of degree at most $D_0$ with
\bb
p_n(0)& = 1,\qquad\max_{j\in\{0,\ldots,n\}}|p_n(j)| \le 1,\\
\max_{j\in\{1,\ldots,n\}}|p_n(j)|& \le \epsilon_0 \coloneqq  e^{-cD_0^2/n}
\ee
for a universal $c>0$ and $D_0 \ge 3$, as holds for all sufficiently large $n$. Its Newton expansion is
\bb
p_n(L_n) = \sum_{s' = 0}^{D_0}a_{s'}\sum_{|S'| = s'}\prod_{j\in S'}\Delta_j, \qquad |a_{s'}| \le 2^{s'}.
\label{sc_newton}
\ee
In particular $\scnorm{p_n(L_n) - \Pi^{(n)}}_\infty \le \epsilon_0$.

\emph{The approximant and its operator-norm error.} Set
\bb
A_n = P p_n(L_n)q_n(K^{(n)}/n)P.
\label{sc_approximant}
\ee
It has finite local output support. Outside $[0,T]$, $|q_n(x)| \le (1 + 2x/T)^{D_1} \le  e^{2D_1x/T}$ for $x \ge 0$. Thus \eqref{sc_reverse} gives
\bb
\scnorm{q_n(K^{(n)}/n)\id_{\{K^{(n)}>nT\}}P}_\infty \le \epsilon_2 \coloneqq  e^{-n[sT - \ell_{U^\dag}(s,2F_0) - 4t_1]/2}.
\ee
Choose $t_1<1/4$. The exponent is then strictly positive. On the complementary spectral interval $q_n$ is bounded by one and differs from $f_n$ by at most $\epsilon_1$. All functions of $K^{(n)}$ commute with $p_n(L_n)$ and $\Pi^{(n)}$. Consequently
\bb
\scnorm{A_n - G_n}_\infty \le 2e^{-n[sF_0 - \ell_U(s,E_1)]/2} + \epsilon_1 + \epsilon_2 + \epsilon_0(1 + \epsilon_2).
\label{sc_error}
\ee
For fixed positive $t_0,t_1$ this is exponentially small.

\emph{Projective-norm estimate.} It remains to show that the exponent of $\scnorm{A_n}_\pi$ can be arbitrarily small; using a local Fock dimension proportional to $n$ at every site would not suffice. We instead use the total photon number on each small subset. There is a constant $A_U$ such that, on any subset $J$ of $v$ sites,
\bb
\scnorm{\left(\sum_{j\in J}K_j\right)^r P^{\rm phys}_{J, \le  L}}_\infty \le [A_U(L + dv + 2r + 2)]^r.
\label{sc_quadratic}
\ee
To justify this, write $\sum_{j\in J}K_j = \sum_{j\in J}c_j^\dag c_j$ where $P^{\rm phys}_{J, \le  L}$ is the total photon-number cutoff on these sites and the $c_j$ are fixed Bogoliubov linear combinations of the physical modes. The coefficient matrix is block diagonal with a fixed squared operator norm $q_U$, so the column $\mathcal{C}_J = (c_j)_{j\in J}$ obeys $\scnorm{\mathcal{C}_J P^{\rm phys}_{J, \le  L}}_\infty^2 \le  q_U(2L + dv)$. Each $c_j$ changes the photon number by at most one. The adjoint row on vectors of photon number at most $L + 1$ therefore has norm at most $\sqrt{q_U(2L + dv + 4)}$, giving $\scnorm{(\sum_{j\in J}K_j)P^{\rm phys}_{J, \le  L}}_\infty \le  q_U(2L + dv + 2)$. One may take $q_U \leq \kappa_S = \|S\|_\infty^2$, since conversion between quadratures and the doubled annihilation--creation vector is unitary. Iterating, with at most two additional photons per power, proves \eqref{sc_quadratic} with $A_U = 2\kappa_S$. The same largest singular value applies to $S^{-1}$, because $S$ is symplectic.

Expand $q_n(x) = \sum_{r = 0}^{D_1}b_rx^r$ and use the multinomial expansion of $(K^{(n)})^r$. Each resulting term, together with a term of \eqref{sc_newton}, has the form
\bb
\frac{r!}{n^r\prod_jm_j!}\, \left(\prod_jK_j^{m_j}\right)\left(\prod_{j\in S'}\Delta_j\right), \qquad\sum_jm_j = r.
\label{sc_localterm}
\ee
It is supported on at most $D_0 + D_1$ sites. All its factors commute and are nonnegative. For $r = 0$ it is a contraction. For $r \geq 1$, positivity of the terms in the multinomial expansion gives the joint spectral inequality
\bb
\frac{r!}{\prod_jm_j!}\prod_jK_j^{m_j} \leq \left(\sum_{j:m_j>0}K_j\right)^r.
\ee
Since the $\Delta$ product is a commuting contraction, \eqref{sc_quadratic} bounds the norm of every compressed term \eqref{sc_localterm} by $\beta_0^r$, with a constant $\beta_0$ depending on $U,F_0,d$ but independent of $n,r,m,S'$, provided $(D_0 + D_1)/n \le 1$. Here the physical total cutoff on the supporting subset is at most $2nF_0$; one may take $\beta_0 = A_U(2F_0 + d + 4)$. All unbounded-operator comparisons are understood as quadratic-form comparisons, followed by finite photon-number compression. The $\Delta_j$ need not preserve physical photon number: their commutation with the $K_j$ supplies the positive-operator comparison.

For clarity, the compression by $P$ does not preserve small support, but its cost across $B:F$ is only polynomial. On a supporting subset $J$ use
\bb
P_{B, \le  L} = \sum_{a = 0}^L P_{B_J, = a}\otimes P_{B_{J^c}, \le  L - a}, \qquad L = \lfloor nF_0\rfloor,
\ee
and the analogous environment identity on both sides of the local term. This produces at most $(L + 1)^4$ terms. The receiver subset has dimension at most $\binom{L + v}{v}$, where $v = |J|$. The elementary finite-dimensional inequality $\scnorm Z_\pi \le (\dim B_J)^2\scnorm Z_\infty$ bounds each term; its remaining factors are local projections of norm at most one. Therefore the projective-norm cost per compressed local term is at most
\bb
(L + 1)^4\binom{L + D_0 + D_1}{D_0 + D_1}^{\!2}\beta_0^r.
\ee
There are $\binom{n + r - 1}{r} \le \binom{n + D_1}{D_1}$ multinomial index tuples. The coefficient sum of \eqref{sc_binomial}, expanded in $x$, is at most $(1 + 2/T)^{D_1} \le 3^{D_1}$. Finally, $\sum_{s' = 0}^{D_0}2^{s'}\binom{n}{s'} \le 2^{D_0}e^{nh_2(D_0/n)}$ when $D_0 \le  n/2$. These estimates give
\bb
\scnorm{A_n}_\pi \le {}&(L + 1)^4 \binom{L + D_0 + D_1}{D_0 + D_1}^{\!2}\binom{n + D_1}{D_1}\\
&\times(3\max\{1,\beta_0\})^{D_1}2^{D_0}e^{nh_2(D_0/n)}.
\label{sc_projectivenorm}
\ee
Put $t = t_0 + t_1$ and recall the thermal entropy function $g$ defined in Eq.~\eqref{thermal_entropy}. It follows that
\bb
\limsup_{n\to\infty}&\frac1n\ln\scnorm{A_n}_\pi\\
\le {}&2t g(F_0/t) + h_2(t_0) + t_0\ln2\\
&+(1 + t_1)h_2\left(\frac{t_1}{1 + t_1}\right) + t_1\ln(3\max\{1,\beta_0\}).
\label{sc_smallrate}
\ee
The right-hand side tends to zero as $t_0,t_1\downarrow0$. Choose them positive and small enough that it is less than $\eta$. Equation~\eqref{sc_error} still has a strictly positive exponential rate. Absorb fixed prefactors by decreasing that rate to obtain \eqref{sc_approx}.
\end{proof}

\subsection{Proof of the photon-number-cutoff strong converse}
\begin{proof}[Proof of Proposition~\ref{strong_sc_thm}]
Choose $E_1>E$ so close that $C(\Phi,E_1)<R$ and then choose $\eta>0$ with $2\eta<R - C(\Phi,E_1)$. Proposition~\ref{sc_approx_prop} gives $\kappa>0$. Choose
\bb
0<a<\min\left\{\kappa,\frac{R - C(\Phi,E_1) - 2\eta}{2}\right\}.
\ee
If a code had $p \ge  e^{-an}$, then $z \le  e^{-n(\kappa - a/2)}\to0$ uniformly over such codes. For large $n$, $1 - 4z>0$, so substituting $\ln M \ge  nR$, $\ln p \ge - an$, and $\ln\Gamma \le  n\eta$ in \eqref{sc_filtercoding} would imply
\bb
C(\Phi,E_1) \ge  R - 2a - 2\eta - o(1),
\ee
contrary to the choice of $a$. This proves exponential decay for all sufficiently large $n$. At $E = 0$, every input is vacuum and $p = 1/M$ directly.
\end{proof}

\subsection{An explicit exponent at every energy}\label{filterquant_app}
The construction above can be made quantitative without an unspecified polynomial-approximation constant. We first give an explicit image polynomial, and then collect all losses in a finite-parameter exponent bound.

\emph{A discrete orthogonal-polynomial filter.} For $0 \leq  j \leq  n$, let $Q_j(x) = Q_j(x;0,0,n)$ be the Hahn polynomial normalised by $Q_j(0) = 1$. Its finite-sum expression and uniform-grid orthogonality are~\cite[Secs.~18.19--18.20]{DLMFHahn}
\bb
Q_j(x)& = \sum_{l = 0}^j(-1)^l \binom jl\binom{j + l}l\frac{\binom xl}{\binom nl},\\
\sum_{x = 0}^n Q_j(x)Q_k(x)& = \frac{\delta_{jk}}{w_j},\qquad w_j = \frac{(2j + 1)(n!)^2}{(n - j)!(n + j + 1)!}.
\label{sc_hahn}
\ee
For integers $0 \leq  D<n$ set
\bb
Z_D& = \frac{(n!)^2}{(n - D - 1)!(n + D + 1)!},\\
p_{n,D}(x)& = \frac{\sum_{j = 0}^D w_jQ_j(x)}{1 - Z_D}.
\label{sc_kernel}
\ee
The weights telescope to $\sum_{j = 0}^D w_j = 1 - Z_D$. Consequently $p_{n,D}(0) = 1$, and orthogonality gives the exact identity
\bb
\sum_{x = 1}^n |p_{n,D}(x)|^2 = \frac{Z_D}{1 - Z_D}.
\label{sc_kernelerror}
\ee
Thus the maximum error away from zero is at most $\sqrt{Z_D/(1 - Z_D)}$. For $D = \lfloor t_0n\rfloor$ with $0<t_0<1$, Stirling's formula yields
\bb
\max_{x\in\{1,\ldots,n\}}|p_{n,D}(x)| & \leq  e^{-nr_0(t_0) + o(n)},\\
r_0(t_0)& = \tfrac12(1 + t_0)\ln(1 + t_0)\\
&\quad + \tfrac12(1 - t_0)\ln(1 - t_0)>0.
\label{sc_norrate}
\ee
For all sufficiently large $n$, this polynomial is bounded by one on the integer grid, so its Newton coefficients obey the same $2^{s'}$ bound as in Eq.~\eqref{sc_newton}. It can therefore replace the image polynomial in Proposition~\ref{sc_approx_prop} without changing the projective-norm estimate. This gives a direct construction of the small-error NOR approximation used here.

\emph{Constants from a chosen Gaussian dilation.} Let $S$ be the symplectic matrix of the centred $d$-mode unitary dilation and put $K = \|S\|_\infty^2 \geq 1$. Both $U$ and $U^\dag$ satisfy the moment bound with
\bb
c = (K - 1)/2,\qquad a_S = (K - K^{-1})/2.
\ee
Indeed, the smallest eigenvalue of $SS^{\mathsf{T}}$ is $K^{-1}$. Write $\ell(s,L)$ for Eq.~\eqref{sc_ell} with these constants and $\kappa_S = K$. All parameters below are scalar and explicitly computable. Choose $E_1>E$, $T>\max\{1,E_1\}$, $F_0>0$, and an admissible $s>0$, as well as $0<t_0<1/2$, $t_1>0$ with $t = t_0 + t_1<1$. Define
\bb
B = {}&2t g(F_0/t) + h_2(t_0) + t_0\ln2\\
&+(1 + t_1)h_2\!\left(\frac{t_1}{1 + t_1}\right) + t_1\ln(3\max\{1,\beta_0\}),\\
\beta_0 = {}&2K(2F_0 + d + 4),\\
\kappa = {}&\min\left\{ \frac{sF_0 - \ell(s,E_1)}2,\; \frac{t_1(E_1 - E)^2}{8T^2},\right.\\
&\hspace{15mm}\left. \frac{sT - \ell(s,2F_0) - 4t_1}{2},\;r_0(t_0)\right\}.
\label{sc_explicitconstants}
\ee
Restrict to choices with $\kappa>0$. The direct and reverse cutoff conditions can always be met by first choosing $F_0$, then $T$, and finally small positive $t_0,t_1$.

\begin{proposition}[(Computable filter exponent)]\label{filterexponent_prop}
For codes at rate at least $R$ supported on the subspace spanned by Fock states with total photon number at most $\lfloor nE\rfloor$, with $E>0$,
\bb
\liminf_{n\to\infty} - \frac1n\ln p_{\rm succ}^{(n)} \geq \mathcal{E}_{\rm filt}(R,E),
\ee
where $[t]_+ \coloneqq \max\{t,0\}$ and, for any fixed Gaussian dilation,
\bb
\mathcal{E}_{\rm filt}(R,E) = \sup_{E_1,s,F_0,T,t_0,t_1} \left[\min\left\{2\kappa,\frac{R - C(\Phi,E_1) - 2B}{2}\right\}\right]_+.
\label{sc_filterexponent}
\ee
The supremum runs over the admissible parameters just specified. It is strictly positive whenever $R>C(\Phi,E)$.
\end{proposition}
\begin{proof}
The construction with Eq.~\eqref{sc_kernel} gives $\|A_n - G_n\|_\infty \leq  e^{-n\kappa + o(n)}$ and $\|A_n\|_\pi \leq  e^{nB + o(n)}$. For any positive $a<\min\{2\kappa,[R - C(\Phi,E_1) - 2B]/2\}$, a code with $p \geq  e^{-an}$ would have $z \leq  e^{-n(\kappa - a/2) + o(n)}\to0$. The filter inequality would then give $C(\Phi,E_1) \geq  R - 2a - 2B - o(1)$, a contradiction. Taking the supremum proves the exponent bound. For positivity, first choose $E_1$ with $C(\Phi,E_1)<R$, fix valid $s,F_0,T$, and then take $t_0,t_1$ small enough that $2B<R - C(\Phi,E_1)$. Every term defining $\kappa$ remains positive.
\end{proof}

The bound also quantifies its approach to capacity. For every fixed channel and $E>0$, there exist $c_E,\delta_E>0$ such that
\bb
\mathcal{E}_{\rm filt}(C(\Phi,E) + \delta,E) \geq  c_E\frac{\delta^3}{\ln(e/\delta)},\qquad 0<\delta<\delta_E.
\label{sc_nearthreshold}
\ee
To see this, concavity and $C(\Phi,0) = 0$ imply $C(\Phi,E + v) - C(\Phi,E) \leq [C(\Phi,E)/E]v$. Take $E_1 - E = \delta/(4\Lambda)$ with $\Lambda = 1 + C(\Phi,E)/E$, and choose fixed $s,F_0,T$ valid uniformly for $E_1 \leq  E + 1$. Set $t_0 = t_1 = b\delta/\ln(e/\delta)$ with a sufficiently small fixed $b>0$. Equation~\eqref{sc_explicitconstants} gives $B = O(b\delta)$, so $2B \leq \delta/4$ for small $\delta$. The energy-polynomial error rate is a positive constant times $\delta^3/\ln(e/\delta)$; the NOR rate is of order $\delta^2/\ln^2(e/\delta)$, while the two tail rates stay bounded away from zero. Equation~\eqref{sc_filterexponent} proves the claim. This lower bound is not asserted to be the optimal scaling near capacity.

The exponent is accessible to numerical optimisation, but conservative: the dilation moment estimates and projective-norm estimates need not be simultaneously tight. Where both apply, $\max\{\mathcal{E}_{\rm filt},\mathcal{E}_{\rm ph}\}$ is a valid lower bound on the success exponent. Determining the optimal exponent is a separate problem.

\subsection{Vanishing photon-number tails}
For a weaker photon-number-tail condition at energy $E$, $\Tr[(\id - P_{n,E})\bar\rho_n] \leq \delta_n$, the proof of Eq.~\eqref{sectorbound} gives, with $B_n = B_n(M_n,E)$,
\bb
p_{\rm succ}^{(n)} \leq \left(\sqrt{B_n} + \sqrt{\delta_n}\right)^2.
\ee
Thus $B_n \leq  e^{-n\kappa}$ and $\delta_n \leq  e^{-n\beta}$ imply a success exponent at least $\min\{\kappa,\beta\}$. A merely vanishing tail yields a strong converse but need not yield exponential decay: mixing a reliable higher-energy code into vacuum with weight $e^{-\sqrt n}$ gives both a tail at most this weight and success at least this weight times the original success. An exponential statement must therefore specify the photon-number-tail decay as well as the rate.

\subsection{Fixed finite-mode channels}\label{filtermultimode_app}
The filter argument needs no explicit expression for $C$. In fact, it also applies to a fixed Gaussian channel with $m_A$ input modes and $m_B$ output modes per use, at its regularised energy-constrained Holevo threshold
\bb
C_{\rm reg}(E) = \sup_{n \ge 1}\frac1n\chi(\Phi^{\otimes n},nE).
\label{sc_regularised}
\ee
Here $E$ is the total photon budget per use of the $m_A$-mode channel. Product ensembles give superadditivity at fixed $E$ and, by time sharing, concavity of $C_{\rm reg}(E)$. Holevo data processing bounds it by $m_A\,g(E/m_A)$, so it is finite and continuous for $E>0$. Equation~\eqref{sc_holevo} therefore holds with $C = C_{\rm reg}$ by definition.

For completeness, the dimension-dependent estimates change as follows. Replace $K_j$ by the transformed total photon number of the $m_A$ input modes at site $j$, and let $\Pi_j$ impose vacuum on its auxiliary inputs as before. The Gaussian moment bound, polynomial construction, and positive-filter lemma are unchanged. The receiver subset dimension becomes $\binom{L + m_Bv}{m_Bv}$, and the first term in \eqref{sc_smallrate} becomes $2m_Bt\,g(F_0/(m_Bt))$, which still tends to zero with $t$. Thus every rate above $C_{\rm reg}(E)$ has exponentially vanishing success for inputs supported on the subspace spanned by Fock states with at most $\lfloor nE\rfloor$ photons in total across the $nm_A$ input modes.

The same projection and vacuum-dilution arguments give
\bb
C_\epsilon(\Phi,E) = C_{\rm reg}\!\left(\frac{E}{1 - \epsilon}\right), \qquad 0 \leq \epsilon<1,
\ee
under the total mean photon budget per use and average error. Here the regularised Holevo expression is the operational vanishing-error capacity under the usual energy constraint~\cite{HolevoBook,HolevoShirokov2006}. For general multimode channels, the converse and rate--error relation therefore hold in terms of the regularised capacity, whose evaluation remains open.

The number of modes is fixed in this statement. Growing memory blocks require uniform control of their normal-mode dilations; Appendix~\ref{memory_sc_app} supplies it for the correlated-noise family considered here.

\section{A computable photon-number converse exponent}\label{photon_sc_app}
\begin{proposition}[(Explicit photon-tail bound)]\label{photon_sc_prop}
For every Gaussian channel in canonical form~\eqref{canonical} and $E \geq 0$, codes supported on the subspace spanned by Fock states with total photon number at most $\lfloor nE\rfloor$ obey
\bb
p_{\rm succ}^{(n)} \leq 4\exp[-n\mathcal{E}_{\rm ph}(R_n,E)],
\label{photon_sc}
\ee
where Eq.~\eqref{photonexponent} gives a computable bound with $\mathcal{E}_{\rm ph}(R,E)>0$ whenever $R>U(\Phi,E)$.
\end{proposition}
In this appendix, $H_n$ denotes the total photon number on the input or output modes, as appropriate. The proof has two independent ingredients: an output photon-number bound that retains the input cutoff, and the unconstrained minimum-output R\'enyi theorem. Squeezing can be used for the latter because no input constraint enters that minimisation. We do not use it to transform the input cutoff.

\subsection{A uniform exponential-moment bound}
Write $y_x = y/\xi$, $y_p = y\xi$, and define
\bb
a_+& = (y_x + 1 - k)/2,\qquad a_- = (y_p + 1 - k)/2,\\
q_s& = 1 - e^{-s},\qquad u_s = \frac{kq_s}{1 - a_+q_s},\\
L_\Phi(s,E)& = -s - \frac12\ln[(1 - a_+q_s)(1 - a_-q_s)]\\
&\quad - (E + 1)\ln(1 - u_s).
\label{mgfbound}
\ee
The domain $\mathcal{S}$ consists of $s>0$ with $q_s(k + a_+)<1$. It contains a right neighbourhood of zero, and all logarithms in Eq.~\eqref{mgfbound} have positive arguments there.

\begin{lemma}[(Output photon-number moments)]\label{mgf_lemma}
For every $n$-mode state $\rho$ supported on the subspace spanned by Fock states with total photon number at most $\lfloor nE\rfloor$ and every $s\in\mathcal{S}$,
\bb
\Tr\big[\Phi^{\otimes n}(\rho)e^{sH_n}\big] \leq  e^{nL_\Phi(s,E)}.
\label{mgfuniform}
\ee
Moreover, $L_\Phi(0,E) = 0$ and $\partial_sL_\Phi(0,E) = N_{\rm out}(E)$.
\end{lemma}

\begin{proof}
Use real heterodyne outcomes $Z\in\mathbb{R}^{2n}$, normalised so that a coherent amplitude $\alpha$ corresponds to $(\sqrt2\,\mathrm{Re}\,\alpha,\sqrt2\,\mathrm{Im}\,\alpha)$. A Gaussian input with covariance matrix $V$ has heterodyne covariance matrix $(V + \id)/2$, consistently with $\hbar = 1$. For an arbitrary state, the radial Husimi identity is
\bb
\mathbb{E}_{Q_\rho}e^{u|Z|^2/2} = \Tr\big[\rho(1 - u)^{-(H_n + n)}\big],\qquad 0 \leq  u<1.
\label{husimiradial}
\ee
It follows by integrating the coherent-state Fock expansion; angular integration removes off-diagonal terms, and each radial integral is a gamma integral. Thus the right-hand side is at most $(1 - u)^{-n(E + 1)}$ under the input cutoff. No independence between input modes is assumed.

Temporarily smooth the output Husimi distribution by adding a centred classical Gaussian of covariance matrix $c\id_{2n}$, with $c = (v - 1)/2$ and $v \geq \max\{1,k - y_p\}$. Its characteristic function is that of the output Wigner distribution multiplied by $e^{-v|\zeta|^2/4}$. The Gaussian channel action shows that this smoothed distribution is exactly the law of
\bb
Z_v&=X^{\oplus n}Z+G,\\
\operatorname{Cov}(G)&=A_v^{\oplus n},\\
A_v&=\frac{Y + (v - k)\id}{2}\geq0 \, ,
\ee
where $Z$ has the input Husimi distribution and $G$ is independent. This identity is valid for all input states, including entangled ones.

Put $t = q_s/(1 + cq_s)$. Gaussian integration over the smoothing noise gives, for $\sigma = \Phi^{\otimes n}(\rho)$,
\bb
\Tr\big[\sigma e^{sH_n}\big] = e^{-ns}(1 + cq_s)^{-n} \mathbb{E} e^{t|Z_v|^2/2}.
\label{smoothingidentity}
\ee
Integration over $G$, conditional on $Z$, gives
\bb
\mathbb{E}[e^{t|Z_v|^2/2}\mid Z] & = \det(\id - tA_v)^{-n/2}\\
&\quad\times\exp\!\left[\frac t2 Z^{\mathsf{T}} \bigl(X^{\mathsf{T}}(\id - tA_v)^{-1}X\bigr)^{\oplus n}Z\right].
\ee
Since $XX^{\mathsf{T}} = k\id$, its quadratic form is at most $u_s|Z|^2/2$. Indeed, $t k/[1 - t\lambda_{\max}(A_v)] = u_s<1$. Apply Eq.~\eqref{husimiradial}. Substitution into Eq.~\eqref{smoothingidentity} cancels the auxiliary smoothing parameter, because
\bb
\det(\id - tA_v) = \frac{(1 - a_+q_s)(1 - a_-q_s)}{(1 + cq_s)^2}.
\ee
The result is Eq.~\eqref{mgfuniform}. These nonnegative integrals can first be interpreted in the extended sense; the finite upper bound proves their integrability. Finally, direct differentiation gives
\bb
\partial_sL_\Phi(0,E) = \frac{a_++a_-}{2} + k(E + 1) - 1 = N_{\rm out}(E).
\ee
\end{proof}

Let $P_{n,F}$ project onto output photon number at most $\lfloor nF\rfloor$. For every $F>N_{\rm out}(E)$, Markov's inequality gives the uniform tail bound
\bb
\Tr\big[(\id - P_{n,F})\Phi^{\otimes n}(\rho)\big]&\leq e^{-nJ_s(F,E)},\\
J_s(F,E)&\coloneqq sF - L_\Phi(s,E) \, .
\label{outputtail}
\ee
The derivative in Lemma~\ref{mgf_lemma} ensures $J_s(F,E)>0$ for sufficiently small positive $s$. For pure loss, Eq.~\eqref{mgfbound} reduces to $L_\Phi(s,E) = E\ln(1 - k + ke^s)$, the binomial generating function; the general formula allows squeezed noise and amplifying or conjugating channels.

\subsection{Decoding bound and exponent}
The established Gaussian majorisation theorem~\cite{GHM2015} implies, for every $\alpha>1$ and every input,
\bb
\|\Phi^{\otimes n}(\rho)\|_\alpha & \leq \exp[-n(\alpha - 1)s_\alpha/\alpha],\\
s_\alpha& = \frac{\ln[(n_0 + 1)^\alpha - n_0^\alpha]}{\alpha - 1},
\label{minrenyi}
\ee
where $\|A\|_\alpha = (\Tr\big[|A|^\alpha\big])^{1/\alpha}$ is the Schatten norm, $n_0 = (k + y - 1)/2$, and $s_\alpha\to S_{\min}$ as $\alpha\downarrow1$. To apply the theorem, write $\Phi = \mathcal{U}_S\circ\Psi\circ\mathcal{U}_S^{-1}$, where $S = \diag(\xi^{-1/2},\xi^{1/2})$ and $\Psi$ has the same transfer matrix and isotropic noise $y\id$. The optimisation in Eq.~\eqref{minrenyi} is unconstrained, so input squeezing is harmless at this step. For $\tau<0$, full input transposition converts the isotropic channel to the covariant one with transfer $\sqrt k\id$ and the same noise; the latter is completely positive since $y \geq  k + 1$. This also establishes Eq.~\eqref{minrenyi} for conjugating tensor powers.

For a decoder $\{\Lambda_m\}$, put $T_m = P_{n,F}\Lambda_mP_{n,F}$. These positive contractions sum to $P_{n,F}$. Its rank obeys
\bb
d_n(F) = \binom{n + \lfloor nF\rfloor}{n} \leq  e^{ng(F)}.
\ee
H\"older's inequality, Eq.~\eqref{minrenyi}, and concavity of $x^{(\alpha - 1)/\alpha}$ give
\bb
p_L& \coloneqq \frac1{M_n}\sum_m\Tr\big[T_m\Phi^{\otimes n}(\rho_m)\big]\\
& \leq  e^{-n(\alpha - 1)s_\alpha/\alpha} \left(\frac{d_n(F)}{M_n}\right)^{(\alpha - 1)/\alpha}\\
& \leq  e^{-nA_\alpha(R_n,F)},\\
A_\alpha(R,F)& = \frac{\alpha - 1}{\alpha}[R - g(F) + s_\alpha].
\ee
In detail, $\|T_m\|_{\alpha/(\alpha - 1)} \leq (\Tr\big[T_m\big])^{(\alpha - 1)/\alpha}$ and $\sum_m\Tr\big[T_m\big] = d_n(F)$. The sector Cauchy--Schwarz bound~\eqref{sectorbound}, now applied at the output, yields
\bb
p_{\rm succ}^{(n)} \leq \left(e^{-nA_\alpha(R_n,F)/2} + e^{-nJ_s(F,E)/2}\right)^2.
\label{finitephoton_sc}
\ee
Consequently Eq.~\eqref{photon_sc} holds with
\bb
\mathcal{E}_{\rm ph}(R,E) = \sup_{\substack{\alpha>1,\ s\in\mathcal{S}\\ F>N_{\rm out}(E)}} \Big[\min\big\{A_\alpha(R,F),J_s(F,E)\big\}\Big]_+.
\label{photonexponent}
\ee
For $R>U(\Phi,E)$, first take $F>N_{\rm out}(E)$ sufficiently close to it, then $\alpha>1$ sufficiently close to one, to make $A_\alpha>0$; next take a small positive $s$ to make $J_s>0$. This proves strict positivity and Proposition~\ref{photon_sc_prop}. The exponent is a lower bound, not an assertion of optimality.

\subsection{Cutoff achievability}\label{cutoffach_app}
For $E>0$, choose an optimal Gaussian input ensemble at energy $E'<E$. Its average input is a centred Gaussian state, whose photon-number generating function is finite near zero. Independent random displacements over $n$ modes therefore give an exponentially small average weight above $\lfloor nE\rfloor$. The usual random-coding construction can be chosen with both vanishing decoding error and vanishing cutoff weight: their expected sum tends to zero. Project each codeword onto the cutoff and replace the discarded weight by vacuum. The gentle measurement bound and Jensen's inequality show that this changes the average decoding success by a quantity tending to zero. Thus every rate below $C(\Phi,E')$ is achievable with strict support. Taking $E'\uparrow E$ proves that the cutoff capacity equals $C(\Phi,E)$ at every energy; at $E = 0$ both vanish.

The gap between $U$ and $C$ at input energies below $N_{\rm th}$ in~\eqref{threshold} reflects the looseness of this computable exponent bound. Proposition~\ref{strong_sc_thm} establishes exponential decay at every rate above $C$ by a separate argument that retains the input-energy constraint in the channel image.

\section{An explicit converse for Gaussian signals}\label{gauss_sc_app}
For a nondegenerate Gaussian channel in canonical form~\eqref{canonical} and $N>0$, a distinct and experimentally natural constraint bounds the energy of the displacement in every codeword. Fix an optimal squeezed-vacuum state $\gamma_* = \gamma_{u_*}$ with covariance matrix $V_* = \diag(u_*,u_*^{-1})$ and seed energy $N_s = (\Tr\big[V_*\big] - 2)/4$. Consider deterministic codewords
\bb
\rho_m = \bigotimes_{j = 1}^n D(d_{mj})\gamma_*D(d_{mj})^\dag, \qquad \frac12\sum_j|d_{mj}|^2 \leq  n(N - N_s).
\label{gausscodes}
\ee
Here $D(d)^\dag R D(d) = R + d$, so displacement by $d$ adds $|d|^2/2$ photons. The displacement vectors may be chosen jointly and need not follow a Gaussian distribution. The decoder is unrestricted. Unlike a mean constraint on arbitrary density operators, Eq.~\eqref{gausscodes} excludes dilution with vacuum.

\begin{proposition}[(Exponential converse for the optimal signal family)]\label{gauss_sc_prop}
For the codes~\eqref{gausscodes}, the function $r_\alpha$ in Eq.~\eqref{renyiradius} satisfies $r_\alpha\to C(\Phi,N)$ as $\alpha\downarrow1$ and
\bb
p_{\rm succ}^{(n)} \leq \exp\!\left[-n\frac{\alpha - 1}{\alpha}(R_n - r_\alpha)\right]
\label{gauss_sc}
\ee
for all $\alpha\in(1,2]$ in its domain. Hence every fixed rate above $C(\Phi,N)$ has a strictly positive exponential converse bound for this signal family.
\end{proposition}

We give explicit formulas in the convention $V_{\rm vac} = \id$. For a positive single-mode covariance matrix $V$ with $\mu_V = \sqrt{\det V}>1$, define
\bb
V^{[a]} = \frac{\coth(a\ell_V/2)}{\mu_V}V, \qquad \ell_V = \ln\frac{\mu_V + 1}{\mu_V - 1}.
\ee
This is the covariance matrix of the normalised $a$th power of its centred Gaussian state, for $a>0$. If $\mu_V=1$, we use the continuous extension $V^{[a]}=V$. The optimal average input has covariance matrix $\diag(z_*,w_*)$, as in Appendix~\ref{radius_app}. Let
\bb
V = XV_*X^{\mathsf{T}} + Y,\quad W = X\diag(z_*,w_*)X^{\mathsf{T}} + Y.
\ee
For $1<\alpha \leq 2$, put $r = \alpha - 1$ and
\bb
K_\alpha& = W^{[r]} - V^{[\alpha]},\\
t_\alpha(V)& = [((\mu_V + 1)/2)^\alpha - ((\mu_V - 1)/2)^\alpha]^{-1},\\
b_r(W)& = [((\mu_W - 1)/2)^{-r} - ((\mu_W + 1)/2)^{-r}]^{-1}.
\ee
For $\mu_V=1$, the corresponding continuous extension is $t_\alpha(V)=1$.
The admissible domain $\mathcal{A}$ consists of these orders with $K_\alpha>0$. It contains a right neighbourhood of $1$, since $W^{[r]}$ diverges as a positive matrix when $r\downarrow0$. Define
\bb
A_\alpha& = \frac1r\ln\frac{t_\alpha(V)b_r(W)}{\sqrt{\det(K_\alpha/2)}},\\
B_\alpha& = \frac{1}{r}X^{\mathsf{T}}K_\alpha^{-1}X,\qquad r_\alpha = A_\alpha + 2(N - N_s)\lambda_{\max}(B_\alpha).
\label{renyiradius}
\ee
The Gaussian Petz--R\'enyi formula~\cite{Seshadreesan2018}, in the same covariance matrix convention, gives
\bb
D_\alpha\Bigl(\Phi\big(D(d)\gamma_*D(d)^\dag\big)\Big\|\sigma_*\Bigr) = A_\alpha + d^{\mathsf{T}}B_\alpha d,
\label{gaussrenyi}
\ee
where $D_\alpha(\rho\|\sigma) = r^{-1}\ln\Tr\big[\rho^\alpha\sigma^{1 - \alpha}\big]$. For product codewords, additivity of this divergence and the per-codeword displacement-energy bound give
\bb
D_\alpha\Big(\Phi^{\otimes n}(\rho_m)\Big\|\sigma_*^{\otimes n}\Big) \leq  nr_\alpha.
\ee
To see its operational consequence directly, append the uniformly distributed message label and compare the resulting classical--quantum state with the product of that label and $\sigma_*^{\otimes n}$. Its divergence is at most $nr_\alpha$, because the exponential of $rD_\alpha$ is the average of the corresponding exponentials for the codewords. The decoder's binary success test has probability $p_{\rm succ}$ under the actual state and $1/M_n$ under the reference. Data processing for Petz divergence at $1<\alpha \leq 2$ yields
\bb
nr_\alpha \geq \frac{\alpha}{\alpha - 1}\ln p_{\rm succ} + \ln M_n,
\ee
which proves Eq.~\eqref{gauss_sc}.

At order $1$, understood as the limit $\alpha\downarrow1$, the displacement coefficient in Eq.~\eqref{gaussrenyi} is $B_1 = 2\diag(h_z,h_w)$, with the derivatives evaluated at $(z_*,w_*)$. The optimal random displacements lie in the eigenspace of its largest eigenvalue; the inactive direction at input energies $N<N_{\rm th}$, with $N_{\rm th}$ defined in~\eqref{threshold}, has a smaller one. Furthermore,
\bb
A_1 = C(\Phi,N) - 2(N - N_s)\lambda_{\max}(B_1).
\ee
Continuity of these formulas for covariance matrices at $1$ therefore gives $r_\alpha\to C(\Phi,N)$. At every $R>C(\Phi,N)$, some $\alpha\in\mathcal{A}$ has $r_\alpha<R$, proving a strictly positive bound
\bb
\mathcal{E}_{\rm G}(R,N) = \sup_{\alpha\in\mathcal{A}} \frac{\alpha - 1}{\alpha}[R - r_\alpha]_+.
\label{exponent}
\ee
This gives a converse exponent for the optimal Gaussian signal family with an unrestricted decoder; it need not coincide with the optimal strong-converse exponent over arbitrary inputs. The result follows from the standard R\'enyi method for classical--quantum coding~\cite{MosonyiOgawa2017}, specialised here to the optimal phase-sensitive signal and its energy budget.

\bibliography{references}

\end{document}